\documentclass[preprint,prd,aps,showpacs,showkeys,nofootinbib]{revtex4}
\usepackage{graphicx}
\usepackage{dcolumn}
\usepackage{bm}
\usepackage{color}
\usepackage[dvipsnames]{xcolor}
\usepackage{amssymb}

\definecolor{light-gray}{gray}{0.88}
\definecolor{dark-gray}{gray}{0.40}

\begin{document}

\title{$Z$ boson decays $Z\rightarrow{{l_i}^{\pm}{l_j}^{\mp}}$ and Higgs boson decays $h\rightarrow{{l_i}^{\pm}{l_j}^{\mp}}$
with lepton flavor violation in a $U(1)$ extension of the MSSM}
\author{Yi-Tong Wang$^{1,2}$\footnote{wyt$\_$991222@163.com}, Shu-Min Zhao$^{1,2}$\footnote{zhaosm@hbu.edu.cn}, Tong-Tong Wang$^{1,2}$, Xi Wang$^{1,2}$, Xin-Xin Long$^{1,2}$, Jiao Ma$^{1,2}$, Tai-Fu Feng$^{1,2,3}$}

\affiliation{$^1$ Department of Physics, Hebei University, Baoding 071002, China}
\affiliation{$^2$ Key Laboratory of High-precision Computation and Application of Quantum Field Theory of Hebei Province, Baoding 071002, China}
\affiliation{$^3$ Department of Physics, Chongqing University, Chongqing 401331, China}
\date{\today}

\begin{abstract}

 $U(1)_X$SSM is the extension of the minimal supersymmetric standard model (MSSM) and its local
 gauge group is $SU(3)_C\times SU(2)_L \times U(1)_Y \times U(1)_X$. We study lepton flavor violating (LFV) decays $Z\rightarrow{{l_i}^{\pm}{l_j}^{\mp}}$($Z\rightarrow e{\mu}$, $Z\rightarrow e{\tau}$, and $Z\rightarrow {\mu}{\tau}$) and $h\rightarrow{{l_i}^{\pm}{l_j}^{\mp}}$($h\rightarrow e{\mu}$, $h\rightarrow e{\tau}$, and $h\rightarrow {\mu}{\tau}$), in this model. In the numerical results, the branching ratios of
$Z\rightarrow{{l_i}^{\pm}{l_j}^{\mp}}$ are from $10^{-9}$ to $10^{-13}$ and the branching ratios of $h\rightarrow{{l_i}^{\pm}{l_j}^{\mp}}$ are from $10^{-3}$ to $10^{-9}$, which can approach the present experimental upper bounds. Based on the latest experimental data, we analyze the influence of different sensitive parameters on the
branching ratio, and make reasonable predictions for future experiments. The main sensitive parameters and LFV sources are the non-diagonal elements corresponding to the initial and final generations of leptons, which can be seen from the numerical analysis.
\end{abstract}

\keywords{lepton flavor violation, $U(1)_X$SSM, new physics}

\maketitle

\section{Introduction}

 Neutrinos have tiny masses and mix with each other, as proved by the neutrino oscillation experiment\cite{IN0,IN1}. This indicates that the lepton flavor symmetry is not
conservative in the neutrino region. The Large Hadron Collider (LHC) detect about 125 GeV new particle\cite{IN2,IN3}, whose properties are close to those of the Higgs boson,
which is very successful for the standard model (SM). The LFV decays are forbidden in the SM. If LFV of charged lepton is detected, it is the direct evidence of new physics(NP).
The LFV decays of Higgs boson and Z boson are of interest, which open a window of detecting NP beyond the SM.

In Table I, we summarize the current limitations and future prospects
of the three modes of Z boson ($Z\rightarrow e{\mu}$, $Z\rightarrow e{\tau}$, and $Z\rightarrow {\mu}{\tau}$)\cite{T7,T3,T8,T9,T4,T5}. For the Large Electron-Positron Collider (LEP), the most stringent upper limit
is to use a data sample of $5\times10^{6}$ Z bosons produced in $e^+e^-$ collisions\cite{T7}. The LHC has already produced many more Z bosons in pp collisions. The upper limit on the branching ratio by the ATLAS experiment corresponds to $7.8\times10^{8}$ Z bosons produced\cite{T7}, significantly more restrictive than that from the LEP experiments.
For future sensitivity of CEPC and FCC-ee circular $e^+e^-$ colliders in assuming $3\times10^{12}$ Z decays\cite{T3}, they are about six orders of magnitude more than LEP experiments. Moreover, at least for the $Z\rightarrow e{\tau}$ and $Z\rightarrow {\mu}{\tau}$, CEPC/FCC-ee could improve on the present LHC (future HL-LHC) bounds up to four(three) orders of magnitude.

For $h\rightarrow{{l_i}^{\pm}{l_j}^{\mp}}$($h\rightarrow e{\mu}$, $h\rightarrow e{\tau}$, and $h\rightarrow {\mu}{\tau}$), due to low energy constraints, $h\rightarrow e{\mu}$ is the way more suppressed than $h\rightarrow e{\tau}$ and $h\rightarrow {\mu}{\tau}$.  Moreover, the LHC search for $h\rightarrow e{\tau}$ and $h\rightarrow {\mu}{\tau}$ themselves, so that the discovery of LFV at the LHC or future leptonic colliders is still an open possibility. After the discovery of the Higgs boson, some future experiments have been proposed to study the properties of the Higgs boson, including two circular lepton colliders(CEPC and FCC-ee) and a linear lepton collider(ILC). In Table II, we summarize the current limitations and future sensitivity on LFV Higgs decays\cite{Z4,Z1,Z2,T6}.

Combining the experimental data provided by ATLAS and CMS, the upper limits on the LFV branching ratios of $Z\rightarrow e{\mu}$, $Z\rightarrow e{\tau}$, $Z\rightarrow {\mu}{\tau}$ and $h\rightarrow e{\mu}$, $h\rightarrow e{\tau}$, $h\rightarrow {\mu}{\tau}$ at 95\%  confidence level (C.L.) are shown in the Table \ref {III}\cite{Z3,IN4,IN5,IN6,IN7}. The LFV decays can easily occur in NP models beyond the SM, for instance the
supersymmetric models and others\cite{Z8,Z46}. Due to the running of the LHC, the LFV decays have recently been discussed within various theoretical frameworks\cite{Z13,Z14,Z15,Z16,Z17,Z18}.
\begin{table}
\caption{ Current upper limits on LFV Z decays from LEP and LHC experiments and future sensitivity from CEPC/FCC-ee}
\begin{tabular}{|c|c|c|c|}
\hline
decay modes & $\hspace{1.0cm}LEP(95\%C.L.)\hspace{0.9cm}$ & $\hspace{1.0cm}LHC(95\%C.L.)\hspace{0.9cm}$ & $\hspace{0.6cm}CEPC/FCC-ee\hspace{0.5cm}$\\
\hline
$Z\rightarrow e{\mu}$ & $1.7\times10^{-6}$\cite{T7} & $7.5\times10^{-7}$\cite{T7} & $10^{-8}-10^{-10}$\cite{T5}\\
\hline
$Z\rightarrow e{\tau}$ & $9.8\times10^{-6}$\cite{T3,T8} & $5.0\times10^{-6}$\cite{T4} & $10^{-9}$\cite{T5}\\
\hline
$Z\rightarrow {\mu}{\tau}$ & $1.2\times10^{-5}$\cite{T3,T9} & $6.5\times10^{-6}$\cite{T4} & $10^{-9}$\cite{T5}\\
\hline
\end{tabular}
\label{I}
\end{table}
\begin{table}
\caption{ Current upper limits and future sensitivity on LFV Higgs decays  }
\begin{tabular}{|c|c|c|c|}
\hline
decay modes & $\hspace{1.0cm}LHC(95\%C.L.)\hspace{0.9cm}$ & $\hspace{0.7cm}CEPC/FCC-ee\hspace{0.7cm}$ & $\hspace{1.7cm}ILC\hspace{1.7cm}$\\
\hline
$h\rightarrow e{\mu}$  & $6.2\times10^{-5}$\cite{Z4,Z1,Z2} & $1.2\times10^{-5}$\cite{T6} & $2.1\times10^{-5}$\cite{T6}\\
\hline
$h\rightarrow e{\tau}$  & $4.7\times10^{-3}$\cite{Z4,Z1,Z2} & $1.6\times10^{-4}$\cite{T6} & $2.4\times10^{-4}$\cite{T6}\\
\hline
$h\rightarrow {\mu}{\tau}$  & $2.5\times10^{-3}$\cite{Z4,Z1,Z2} & $1.4\times10^{-4}$\cite{T6} & $2.3\times10^{-4}$\cite{T6}\\
\hline
\end{tabular}
\label{II}
\end{table}

 $U(1)_X$SSM is the extension of the MSSM and its local gauge group is $SU(3)_C\times SU(2)_L \times U(1)_Y \times U(1)_X$\cite{Sarah1,Sarah2,Sarah3}.
To obtain this model, three singlet new Higgs superfields and right-handed neutrinos are added to MSSM. In this work, we  analyze the LFV decays  $Z\rightarrow{{l_i}^{\pm}{l_j}^{\mp}}$($Z\rightarrow e{\mu}$, $Z\rightarrow e{\tau}$, and $Z\rightarrow {\mu}{\tau}$)
and $h\rightarrow{{l_i}^{\pm}{l_j}^{\mp}}$($h\rightarrow e{\mu}$, $h\rightarrow e{\tau}$, and $h\rightarrow {\mu}{\tau}$) within the $U(1)_X$SSM model.
Compared with the MSSM, the neutrino masses in the $U(1)_X$SSM are not zero. These new sources enlarge the LFV processes via loop contributions. Therefore, the expected experimental results for the LFV processes may be obtained in the near future.

 In our previous work, we study the LFV decays $l_j\rightarrow{l_i\gamma}$ in the $U(1)_X$SSM\cite{T1}. The numerical
results show that the present experimental limits for the branching ratio of $l_j\rightarrow{l_i\gamma}$ constrain the parameter space of the $U(1)_X$SSM most strictly. In this work, considering the constraint of the present
experimental limits on the branching ratio of $l_j\rightarrow{l_i\gamma}$, we give the influence of slepton flavor mixing parameters on the branching ratios of $Z\rightarrow{{l_i}^{\pm}{l_j}^{\mp}}$ , $h\rightarrow{{l_i}^{\pm}{l_j}^{\mp}}$ and $l_j\rightarrow{l_i\gamma}$ in the $U(1)_X$SSM.

 We work mainly on the following aspects. In Section II, We briefly introduce the main content of $U(1)_X$SSM, including its superpotential,
the general soft SUSY-breaking terms, needed mass matrices and couplings. Section III and IV are devoted to the decays $Z\rightarrow{{l_i}^{\pm}{l_j}^{\mp}}$ and
$h\rightarrow{{l_i}^{\pm}{l_j}^{\mp}}$ with lepton flavor violation in the $U(1)_X$SSM.
In Section V, we give the corresponding parameters and numerical analysis. The discussion and conclusion are given in
Section VI. The appendix introduces some specific forms of coupling coefficient that we need.
\begin{table}
\caption{ The upper limits on the LFV branching ratios of $Z\rightarrow{{l_i}^{\pm}{l_j}^{\mp}}$ and $h\rightarrow{{l_i}^{\pm}{l_j}^{\mp}}$}
\begin{tabular}{|c|c|c|}
\hline
decay modes & Z boson & Higgs boson \\
\hline
$\hspace{1.1cm}e{\mu}\hspace{1.1cm}$ & $\hspace{1.1cm}7.5\times10^{-7}\hspace{1.1cm}$ & $\hspace{1.1cm}6.2\times10^{-5}\hspace{1.1cm}$ \\
\hline
$\hspace{1.1cm}e{\tau}\hspace{1.1cm}$ & $\hspace{1.1cm}9.8\times10^{-6}\hspace{1.1cm}$ & $\hspace{1.1cm}4.7\times10^{-3}\hspace{1.1cm}$  \\
\hline
$\hspace{1.1cm}{\mu}{\tau}\hspace{1.1cm}$ & $\hspace{1.1cm}1.2\times10^{-5}\hspace{1.1cm}$ & $\hspace{1.1cm}2.5\times10^{-3}\hspace{1.1cm}$   \\
\hline
\end{tabular}
\label{III}
\end{table}

\section{The main content of $U(1)_X$SSM}
$U(1)_X$SSM is the U(1) extension of MSSM, and the local gauge group is $SU(3)_C\otimes
SU(2)_L \otimes U(1)_Y\otimes U(1)_X$\cite{UU1,UU3,UU4,T1}. In order to obtain $U(1)_X$SSM, MSSM has added new superfields including:
right-handed neutrinos $\hat{\nu}_i$ and three Higgs singlets $\hat{\eta},~\hat{\bar{\eta}},~\hat{S}$.
Through the seesaw mechanism, light neutrinos obtain tiny mass at the tree level.
The neutral CP-even parts of $H_u,~ H_d,~\eta,~\bar{\eta}$ and $S$ mix together, forming $5\times5 $ mass squared matrix.
The loop corrections to the lightest CP-even Higgs are important because we need it to get the Higgs mass of 125.1 GeV\cite{LCTHiggs1,LCTHiggs2}.
The sneutrinos are disparted into CP-even sneutrinos and CP-odd sneutrinos, and their mass squared matrixes are both extended to $6\times6$.

 The superpotential in $U(1)_X$SSM is expressed as:
\begin{eqnarray}
&&W=l_W\hat{S}+\mu\hat{H}_u\hat{H}_d+M_S\hat{S}\hat{S}-Y_d\hat{d}\hat{q}\hat{H}_d-Y_e\hat{e}\hat{l}\hat{H}_d+\lambda_H\hat{S}\hat{H}_u\hat{H}_d
\nonumber\\&&~~~~~~+\lambda_C\hat{S}\hat{\eta}\hat{\bar{\eta}}+\frac{\kappa}{3}\hat{S}\hat{S}\hat{S}+Y_u\hat{u}\hat{q}\hat{H}_u+Y_X\hat{\nu}\hat{\bar{\eta}}\hat{\nu}
+Y_\nu\hat{\nu}\hat{l}\hat{H}_u.
\end{eqnarray}

 The specific explicit expressions of two Higgs doublets are as follows:
\begin{eqnarray}
&&H_{u}=\left(\begin{array}{c}H_{u}^+\\{1\over\sqrt{2}}\Big(v_{u}+H_{u}^0+iP_{u}^0\Big)\end{array}\right),
~~~~~~
H_{d}=\left(\begin{array}{c}{1\over\sqrt{2}}\Big(v_{d}+H_{d}^0+iP_{d}^0\Big)\\H_{d}^-\end{array}\right).
\end{eqnarray}

 The three Higgs singlets are represented by:
\begin{eqnarray}
&&\eta={1\over\sqrt{2}}\Big(v_{\eta}+\phi_{\eta}^0+iP_{\eta}^0\Big),~~~~~~~~~~~~~~~
\bar{\eta}={1\over\sqrt{2}}\Big(v_{\bar{\eta}}+\phi_{\bar{\eta}}^0+iP_{\bar{\eta}}^0\Big),\nonumber\\&&
\hspace{3.0cm}S={1\over\sqrt{2}}\Big(v_{S}+\phi_{S}^0+iP_{S}^0\Big).
\end{eqnarray}

Here,  $v_u,~v_d,~v_\eta$,~ $v_{\bar\eta}$ and $v_S$ are the corresponding vacuum expectation values(VEVs) of the Higgs superfields $H_u$, $H_d$, $\eta$, $\bar{\eta}$ and $S$.
Two angles are defined as $\tan\beta=v_u/v_d$ and $\tan\beta_\eta=v_{\bar{\eta}}/v_{\eta}$.
The definition of ${\widetilde{\nu}}_{L}$ and ${\widetilde{\nu}}_{R}$ is:
\begin{eqnarray}
&&\widetilde{\nu}_{L}= \frac{1}{\sqrt{2} } {\phi}_{l}+\frac{i}{\sqrt{2} } {\sigma}_{l},~~~~~~~~~~~~~~
\widetilde{\nu}_{R}= \frac{1}{\sqrt{2} } {\phi}_{R}+\frac{i}{\sqrt{2} } {\sigma}_{R}.
\end{eqnarray}

 The soft SUSY breaking terms of $U(1)_X$SSM is:
\begin{eqnarray}
&&\mathcal{L}_{soft}=\mathcal{L}_{soft}^{MSSM}-B_SS^2-L_SS-\frac{T_\kappa}{3}S^3-T_{\lambda_C}S\eta\bar{\eta}
+\epsilon_{ij}T_{\lambda_H}SH_d^iH_u^j\nonumber\\&&\hspace{1.5cm}
-T_X^{IJ}\bar{\eta}\tilde{\nu}_R^{*I}\tilde{\nu}_R^{*J}
+\epsilon_{ij}T^{IJ}_{\nu}H_u^i\tilde{\nu}_R^{I*}\tilde{l}_j^J
-m_{\eta}^2|\eta|^2-m_{\bar{\eta}}^2|\bar{\eta}|^2-m_S^2S^2\nonumber\\&&\hspace{1.5cm}
-(m_{\tilde{\nu}_R}^2)^{IJ}\tilde{\nu}_R^{I*}\tilde{\nu}_R^{J}
-\frac{1}{2}\Big(M_X\lambda^2_{\tilde{X}}+2M_{BB^\prime}\lambda_{\tilde{B}}\lambda_{\tilde{X}}\Big)+h.c~~.
\end{eqnarray}

 The Table \ref {IV} shows the particle content and charge distribution of $U(1)_X$SSM.
We have shown that $U(1)_X$SSM is anomaly free in previous work\cite{UU4}.
The two Abelian groups $U(1)_Y$ and $U(1)_X$ in $U(1)_X$SSM cause a new effect: the gauge kinetic mixing.
This effect can be induced by RGEs.

\begin{table}
\caption{ The superfields in $U(1)_X$SSM}
\begin{tabular}{|c|c|c|c|c|c|c|c|c|c|c|c|}
\hline
Superfields & $\hspace{0.1cm}\hat{q}_i\hspace{0.1cm}$ & $\hat{u}^c_i$ & $\hspace{0.2cm}\hat{d}^c_i\hspace{0.2cm}$ & $\hat{l}_i$ & $\hspace{0.2cm}\hat{e}^c_i\hspace{0.2cm}$ & $\hat{\nu}_i$ & $\hspace{0.1cm}\hat{H}_u\hspace{0.1cm}$ & $\hat{H}_d$ & $\hspace{0.2cm}\hat{\eta}\hspace{0.2cm}$ & $\hspace{0.2cm}\hat{\bar{\eta}}\hspace{0.2cm}$ & $\hspace{0.2cm}\hat{S}\hspace{0.2cm}$ \\
\hline
$SU(3)_C$ & 3 & $\bar{3}$ & $\bar{3}$ & 1 & 1 & 1 & 1 & 1 & 1 & 1 & 1  \\
\hline
$SU(2)_L$ & 2 & 1 & 1 & 2 & 1 & 1 & 2 & 2 & 1 & 1 & 1  \\
\hline
$U(1)_Y$ & 1/6 & -2/3 & 1/3 & -1/2 & 1 & 0 & 1/2 & -1/2 & 0 & 0 & 0  \\
\hline
$U(1)_X$ & 0 & -1/2 & 1/2 & 0 & 1/2 & -1/2 & 1/2 & -1/2 & -1 & 1 & 0  \\
\hline
\end{tabular}
\label{IV}
\end{table}

The general form of the covariant derivative of $U(1)_X$SSM can be found in Refs.\cite{UMSSM5,B-L1,B-L2,gaugemass}. In $U(1)_X$SSM, the gauge bosons $A^{X}_\mu,~A^Y_\mu$ and $V^3_\mu$ mix together at the tree level. The mass matrix
in the basis $(A^Y_\mu, V^3_\mu, A^{X}_\mu)$ can be found in Ref.\cite{UU4}. We use two mixing angles $\theta_{W}$ and $\theta_{W}'$ to obtain mass eigenvalues of the matrix. $\theta_{W}$ is the Weinberg angle and $\theta_{W}'$ is the new mixing angle. The new mixing angle is defined as
\begin{eqnarray}
\sin^2\theta_{W}'\!=\!\frac{1}{2}\!-\!\frac{[(g_{{YX}}+g_{X})^2-g_{1}^2-g_{2}^2]v^2+
4g_{X}^2\xi^2}{2\sqrt{[(g_{{YX}}+g_{X})^2+g_{1}^2+g_{2}^2]^2v^4\!+\!8g_{X}^2[(g_{{YX}}+g_{X})^2\!-\!g_{1}^2\!-\!g_{2}^2]v^2\xi^2\!+\!16g_{X}^4\xi^4}}.
\end{eqnarray}
Here, $v^2=v_u^2+v_d^2$ and $\xi^2=v_\eta^2+v_{\bar{\eta}}^2$.

 The new mixing angle appears in the couplings involving $Z$ and $Z^{\prime}$.
The exact eigenvalues are calculated
\begin{eqnarray}
&&m_\gamma^2=0,\nonumber\\
&&m_{Z,{Z^{'}}}^2=\frac{1}{8}\Big([g_{1}^2+g_2^2+(g_{{YX}}+g_{X})^2]v^2+4g_{X}^2\xi^2 \nonumber\\
&&\hspace{1.1cm}\mp\sqrt{[g_{1}^2+g_{2}^2+(g_{{YX}}+g_{X})^2]^2v^4\!+\!8[(g_{{YX}}+g_{X})^2\!-\!g_{1}^2\!-\!
g_{2}^2]g_{X}^2v^2\xi^2\!+\!16g_{X}^4\xi^4}\Big).
\end{eqnarray}

The mass squared matrix for CP-even Higgs $({\phi}_{d}, {\phi}_{u}, {\phi}_{\eta}, {\phi}_{\overline{\eta}}, {\phi}_{s})$ reads
\begin{eqnarray}
M^2_{h} = \left(
\begin{array}{ccccc}
m_{{\phi}_{d}{\phi}_{d}} &m_{{\phi}_{u}{\phi}_{d}} &m_{{\phi}_{\eta}{\phi}_{d}} &m_{{\phi}_{\overline{\eta}}{\phi}_{d}} &m_{{\phi}_{s}{\phi}_{d}}\\
m_{{\phi}_{d}{\phi}_{u}} &m_{{\phi}_{u}{\phi}_{u}} &m_{{\phi}_{\eta}{\phi}_{u}} &m_{{\phi}_{\overline{\eta}}{\phi}_{u}} &m_{{\phi}_{s}{\phi}_{u}}\\
m_{{\phi}_{d}{\phi}_{\eta}} &m_{{\phi}_{u}{\phi}_{\eta}} &m_{{\phi}_{\eta}{\phi}_{\eta}} &m_{{\phi}_{\overline{\eta}}{\phi}_{\eta}} &m_{{\phi}_{s}{\phi}_{\eta}}\\
m_{{\phi}_{d}{\phi}_{\overline{\eta}}} &m_{{\phi}_{u}{\phi}_{\overline{\eta}}} &m_{{\phi}_{\eta}{\phi}_{\overline{\eta}}} &m_{{\phi}_{\overline{\eta}}{\phi}_{\overline{\eta}}} &m_{{\phi}_{s}{\phi}_{\overline{\eta}}}\\
m_{{\phi}_{d}{\phi}_{s}} &m_{{\phi}_{u}{\phi}_{s}} &m_{{\phi}_{\eta}{\phi}_{s}} &m_{{\phi}_{\overline{\eta}}{\phi}_{s}} &m_{{\phi}_{s}{\phi}_{s}}\end{array}
\right),\label{Rsneu}
 \end{eqnarray}
\begin{eqnarray}
&&m_{\phi_{d}\phi_{d}}= m_{H_d}^2+  |\mu|^2
 +\frac{1}{8} \Big( [g_{1}^{2}+(g_{X}+g_{YX})^{2}+g_2^2] (3 v_{d}^{2}  - v_{u}^{2})\nonumber \\
&&\hspace{1.5cm}+2 (g_{Y X} g_{X}+g_X^2) ( v_{\eta}^{2}- v_{\bar{\eta}}^{2})\Big)+ \sqrt{2} v_S \mu {\lambda}_{H}+\frac{1}{2} (v_{u}^{2} + v_S^{2})|{\lambda}_{H}|^2,
\\&&m_{\phi_{d}\phi_{u}} = -\frac{1}{4} \Big(g_{2}^{2} + (g_{Y X} + g_{X})^{2} + g_1^{2}\Big)v_d v_u
 + |{\lambda}_{H}|^2 v_d v_u - {\lambda}_{H} l_W \nonumber \\
&&\hspace{1.5cm}-\frac{1}{2}{\lambda}_{H} (v_{\eta} v_{\bar{\eta}} {\lambda}_{C}  + v_S^{2} \kappa )
 - B_{\mu}- \sqrt{2} v_S (\frac{1}{2}T_{{\lambda}_{H}}  + M_S {\lambda}_{H} ),
\\ &&m_{\phi_{u}\phi_{u}} = m_{H_u}^2+ |\mu|^2+\frac{1}{8} \Big( [g_{1}^{2}+(g_{X}+g_{YX})^{2}+g_2^2] (3 v_{u}^{2}  - v_{d}^{2})\nonumber \\
&&\hspace{1.5cm}+2 (g_{Y X} g_{X}+g_X^2) ( v_{\bar{\eta}}^{2}-v_{\eta}^{2})\Big)
 +  \sqrt{2} v_S\mu {\lambda}_{H}   + \frac{1}{2}(v_{d}^{2} + v_S^{2})|{\lambda}_{H}|^2,
\\&&m_{\phi_{d}\phi_{\eta}} = \frac{1}{2}g_{X} (g_{Y X} + g_{X})v_d v_{\eta}
  -\frac{1}{2} v_u v_{\bar{\eta}} {\lambda}_{H} {\lambda}_{C} ,
\\&&m_{\phi_{u}\phi_{\eta}} = -\frac{1}{2}g_{X} (g_{Y X} + g_{X})v_u v_{\eta}
-\frac{1}{2} v_d v_{\bar{\eta}} {\lambda}_{H} {\lambda}_{C},
\\&&m_{\phi_{\eta}\phi_{\eta}} = m_{\eta}^2 +\frac{1}{4} \Big((g_{Y X} g_{X}+g_X^2) ( v_{d}^{2}
- v_{u}^{2})+2g_{X}^{2}
( 3 v_{\eta}^{2}-v_{\bar{\eta}}^{2})\Big)+\frac{|{\lambda}_{C}|^2}{2} (v_{\bar{\eta}}^{2} + v_S^{2}),
\\&&m_{\phi_{d}\phi_{\bar{\eta}}} = -\frac{1}{2}g_{X} (g_{Y X} + g_{X})v_d v_{\bar{\eta}}
  -\frac{1}{2} v_u v_{\eta} {\lambda}_{H} {\lambda}_{C},
\\&&m_{\phi_{u}\phi_{\bar{\eta}}} = \frac{1}{2}g_{X} (g_{Y X} + g_{X})v_u v_{\bar{\eta}}  -\frac{1}{2} v_d v_{\eta} {\lambda}_{H} {\lambda}_{C},
\\&&m_{\phi_{\eta}\phi_{\bar{\eta}}} = - g_{X}^{2}v_{\eta} v_{\bar{\eta}}+\frac{1}{2}(2 l_W  - {\lambda}_{H} v_d v_u )
{\lambda}_{C} + |{\lambda}_{C}|^2 v_{\eta} v_{\bar{\eta}}\nonumber \\
&&\hspace{1.5cm} + \frac{1}{\sqrt{2}} v_S (2 M_S {\lambda}_{C} + T_{{\lambda}_{C}}) + \frac{1}{2} v_S^{2} {\lambda}_{C} \kappa,
\\&&m_{\phi_{\bar{\eta}}\phi_{\bar{\eta}}} = m_{\bar{\eta}}^2+\frac{1}{4} \Big((g_{Y X} g_{X}+g_X^2)
 ( v_{u}^{2}- v_{d}^{2})+2g_{X}^{2}( 3 v_{\bar{\eta}}^{2}-v_{\eta}^{2})\Big)+\frac{|{\lambda}_{C}|^2 }{2} \Big(v_{\eta}^{2} + v_S^{2}\Big),
\\&&m_{\phi_{d}{\phi}_{s}} = \Big({\lambda}_{H} v_d v_S  + \sqrt{2} v_d \mu  -  v_u ( \kappa v_S
 + \sqrt{2} M_S )\Big){\lambda}_{H} - \frac{1}{\sqrt{2}}v_u T_{{\lambda}_{H}},
\\&&m_{\phi_{u}{\phi}_{s}} =  \Big( {\lambda}_{H} v_u v_S  + \sqrt{2} v_u \mu
-v_d (\kappa v_S  + \sqrt{2} M_S )\Big){\lambda}_{H}
- \frac{1}{\sqrt{2}}  v_dT_{{\lambda}_{H}},
\\&&m_{\phi_{\eta}{\phi}_{s}} = \Big( {\lambda}_{C} v_{\eta} v_S  + v_{\bar{\eta}} (\kappa v_S
 + \sqrt{2} M_S )\Big){\lambda}_{C}  +\frac{1}{\sqrt{2}}v_{\bar{\eta}} T_{{\lambda}_{C}},
\\&&m_{\phi_{\bar{\eta}}{\phi}_{s}} =\Big( {\lambda}_{C} v_{\bar{\eta}} v_S  + v_{\eta}(\kappa v_S
 + \sqrt{2} M_S )\Big){\lambda}_{C} + \frac{1}{\sqrt{2}}v_{\eta}
 T_{{\lambda}_{C}},
\\&&m_{{\phi}_{s}{\phi}_{s}} = m^2_{S}+ \Big(2 l_W  + 3v_S (\kappa v_S  + 2\sqrt{2} M_S )
+ {\lambda}_{C} v_{\eta} v_{\bar{\eta}}  - {\lambda}_{H} v_d v_u \Big)\kappa\nonumber \\
&&\hspace{1.5cm} +\frac{1}{2}|{\lambda}_{C}|^2 \xi^2+\frac{1}{2}|{\lambda}_{H}|^2 v^{2}
  +2 {B_{S}}  + 4 |M_S|^2   + \sqrt{2} v_S T_{\kappa}.
\end{eqnarray}

This matrix is diagonalized by $Z^H$:
\begin{eqnarray}
Z^H m^2_{h} Z^{H,\dagger} = m^{dia}_{2,h},
 \end{eqnarray}
with
\begin{eqnarray}
{\phi}_{d}=\sum\limits_{j} Z^H_{j1} h_j,~~~{\phi}_{u}=\sum\limits_{j} Z^H_{j2} h_j,~~~{\phi}_{\eta}=\sum\limits_{j}Z^H_{j3}h_j,\nonumber\\
{\phi}_{\overline{\eta}}=\sum\limits_{j} Z^H_{j4} h_j,~~~{\phi}_{s}=\sum\limits_{j} Z^H_{j5} h_j.
 \end{eqnarray}
Other mass matrices can be found in Refs.\cite{UU1,T1}.

Here, we show some of the couplings that we need in the $U(1)_X$SSM. We deduce the vertexes of $Z-\tilde{e}_i-\tilde{e}^{*}_j$
\begin{eqnarray}
&&\mathcal{L}_{Z\tilde{e}\tilde{e}^{*}}=\frac{1}{2}\tilde{e}^{*}_j
\Big[(g_2\cos\theta_W\cos\theta_W^\prime\!-\!g_1\cos\theta_W^\prime\sin\theta_W
+g_{YX}\sin\theta_W^\prime)\sum_{a=1}^3Z_{i,a}^{E,*}Z_{j,a}^E\nonumber\\
&&\hspace{1.5cm}+\Big((2g_{YX}+g_X)\sin\theta_W^\prime\!-\!2g_1\cos\theta_W^\prime\sin\theta_W\Big)
\sum_{a=1}^3Z_{i,3+a}^{E,*}Z_{j,3+a}^E\Big](p^\mu_{i}\!-\!p^\mu_j)\tilde{e}_iZ_{\mu}.
\end{eqnarray}

We also deduce the vertexes of $\bar{l}_i-\chi_j^--\tilde{\nu}^R_k(\tilde{\nu}^I_k)$
\begin{eqnarray}
&&\mathcal{L}_{\bar{l}\chi^-\tilde{\nu}^R}=\frac{1}{\sqrt{2}}\bar{l}_i\Big\{U^*_{j2}Z^{R*}_{ki}Y_l^iP_L
-g_2V_{j1}Z^{R*}_{ki}P_R\Big\}\chi_j^-\tilde{\nu}^R_k,\nonumber\\
&&\mathcal{L}_{\bar{l}\chi^-\tilde{\nu}^I}=\frac{i}{\sqrt{2}}\bar{l}_i\Big\{U^*_{j2}Z^{I*}_{ki}Y_l^iP_L
-g_2V_{j1}Z^{I*}_{ki}P_R\Big\}\chi_j^-\tilde{\nu}^I_k.
\end{eqnarray}

The vertexes of $\bar{\chi}_i^0-l_j-\tilde{e}_k$ are
\begin{eqnarray}
&&\mathcal{L}_{\bar{\chi}_i^0l\tilde{e}}=\bar{\chi}_i^0\Big\{\Big(\frac{1}{\sqrt{2}}(g_1N^*_{i1}+g_2N^*_{i2}+g_{YX}N^*_{i5})Z^E_{kj}-N^*_{i3}Y_l^jZ^E_{k3+j}\Big)P_L\nonumber\\
&&\hspace{1.5cm}-\Big[\frac{1}{\sqrt{2}}(2g_1N_{i1}+(2g_{YX}+g_X)N_{i5})Z^E_{k3+a}+Y_l^jZ^E_{kj}N_{i3}\Big]P_R\Big\}\l_j\tilde{e}_k.
\end{eqnarray}

To save space in the text, the remaining vertexes can be found in the Refs.\cite{UU3,UU4,UU2,UU5}.

\section{Z boson decays $Z\rightarrow{{l_i}^{\pm}{l_j}^{\mp}}$}
\begin{figure}[h]
\setlength{\unitlength}{5.0mm}
\centering
\includegraphics[width=5.0in]{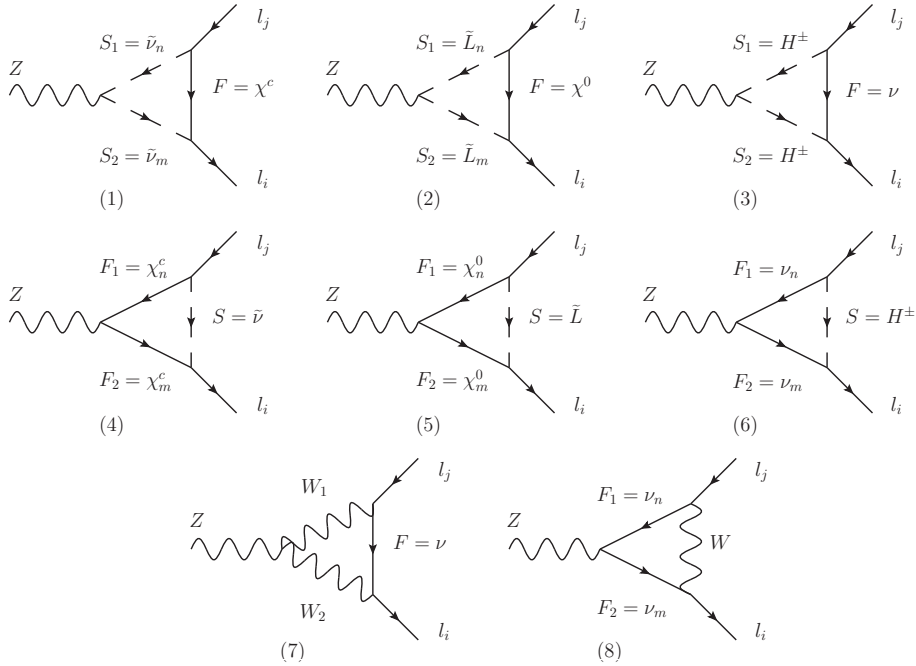}
\caption{ Feynman diagrams for the $Z\rightarrow{{l_i}^{\pm}{l_j}^{\mp}}$ processes in the $U(1)_X$SSM. F represents Dirac (Majorana) fermion, S represents scalar boson, and W represents the W boson.}\label{N1}
\end{figure}

\begin{figure}[h]
\setlength{\unitlength}{5.0mm}
\centering
\includegraphics[width=5.0in]{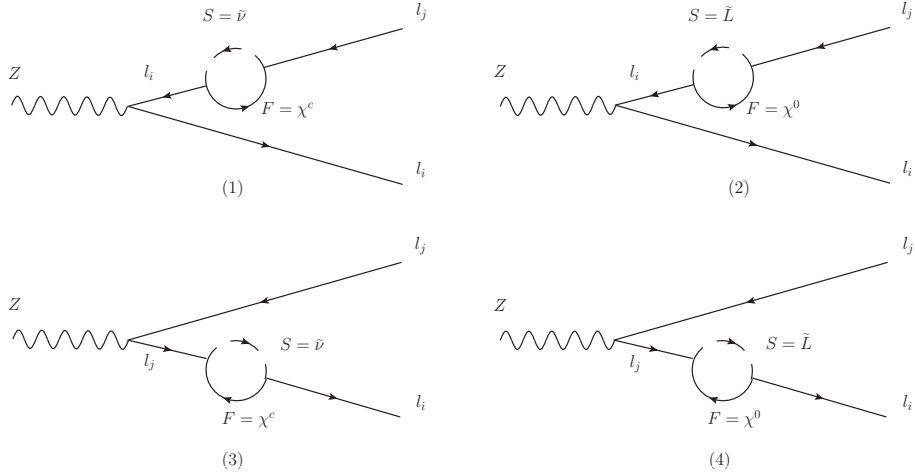}
\caption{ Feynman diagrams for the processes $Z\rightarrow{{l_i}^{\pm}{l_j}^{\mp}}$ in the $U(1)_X$SSM, which denote self-energy diagrams contributing to $Z\rightarrow{{l_i}^{\pm}{l_j}^{\mp}}$ from loops.}\label{N2}
\end{figure}

In this section, we analyze the LFV processes $Z\rightarrow{{l_i}^{\pm}{l_j}^{\mp}}$. The corresponding Feynman diagrams can be depicted by Fig.\ref {N1} and Fig.\ref {N2}, and the corresponding effective amplitudes can be written as\cite{D33,D34,D35}
\begin{eqnarray}
\mathcal{M_\mu}=\overline{l}_i\gamma_\mu(F_LP_L+F_RP_R){l_j},
\end{eqnarray}
with
\begin{eqnarray}
F^Z_{L,R}=F_{L,R}(A)+F_{L,R}(W)+F_{L,R}(B).
\end{eqnarray}

The coefficients $F_{L,R}$ can be obtained from the amplitudes of the Feynman diagrams. $F_{L,R}(A)$ correspond to Fig.\ref{N1}(1) to Fig.\ref {N1}(6),
and stand for the contributions from chargino-sneutrino, neutralino-slepton, neutrino-charged Higgs;
$F_{L,R}(W)$ correspond to Fig.\ref{N1}(7) and Fig.\ref {N1}(8), and stand for the contributions from W-neutrino due to three light neutrinos and three heavy neutrinos mixing together.

 The contributions obtained from Fig.\ref{N1}(1)-Fig.\ref{N1}(6) are expressed by $F_{L,R}(A)=F^{\alpha}_{L,R}(A)(\alpha=1...6)$. The specific forms are as follows:
\begin{eqnarray}
&&F^{(1,2,3)}_L(A)=\frac{i}{2}H^{S_2F\overline{l}_i}_RH^{Z{S_1}S^{\ast}_2}H^{S^{\ast}_1{l_j}\overline{F}}_LG_2(x_F,x_{S_1},x_{S_2}),\nonumber\\
&&F^{(4,5,6)}_L(A)=\frac{i}{2}\Big[\frac{2m_{F_1}m_{F_2}}{m^2_{W}}H^{SF_2\overline{l}_i}_RH^{ZF_1\overline{F}_2}_LH^{S^{\ast}{l_j}\overline{F}_1}_LG_1(x_S,x_{F_1},x_{F_2})\nonumber\\
&&\hspace{1.6cm}-H^{SF_2\overline{l}_i}_RH^{ZF_1\overline{F}_2}_RH^{S^{\ast}{l_j}\overline{F}_1}_LG_2(x_S,x_{F_1},x_{F_2})\Big],\nonumber\\
&&F^{\alpha}_R(A)=F^{\alpha}_L(A)|_{L\leftrightarrow{R}},~~~\alpha=1...6.
\end{eqnarray}

Here, $x_i=m_i^2/m^2_{W}$ with $m_i$ representing the mass of the corresponding particle, and $m_{W}$ representing the energy scale of the NP. Their specific expressions are given in the Appendix. For Fig.\ref{N1}(1), $S_1$ and $S_2$ represent a CP-even scalar neutrino and a CP-odd scalar neutrino. F represents chargino. $H^{S_2F\overline{l}_i}_R$ is the right-handed coupling of the vertex $\tilde{\nu}^{I(R)}-\chi^\pm-\overline{l}_i$. $H^{Z{S_1}S^{\ast}_2}$ is the coupling of $\tilde{\nu}^{R(I)}-Z-\tilde{\nu}^{I(R)}$. $H^{S^{\ast}_1{l_j}\overline{F}}_L$ is the left-handed coupling of the vertex $\tilde{\nu}^{R(I)}-\overline{\chi}^\pm-l_j$. The concrete forms of $H^{S_2F\overline{l}_i}_R$, $H^{Z{S_1}S^{\ast}_2}$ and $H^{S^{\ast}_1{l_j}\overline{F}}_L$ are shown as Eq.(\ref{A1}) in the Appendix. For Fig.\ref{N1}(2), $S_1$ and $S_2$ represent scalar lepton $\tilde{L}$, and F denotes neutralino $\chi^0$. The couplings $H^{\tilde{L}_n\chi^0\overline{l}_i}_R,~H^{Z{\tilde{L}_m}\tilde{L}_n^*}$ and $H^{\tilde{L}_m^*{l_j}\overline{\chi}^0}_L$ are in Eq.(\ref{A2}) of the Appendix. For Fig.\ref{N1}(3), $S_1$ and $S_2$ represent charged Higgs $H^\pm$, and F denotes neutrino $\nu$. The couplings $H^{H^\pm\nu\overline{l}_i}_R,~H^{ZH^\pm H^\pm}$ and $H^{H^\pm{l_j}\overline{\nu}}_L$ are in Eq.(\ref{A3}) of the Appendix.

For Fig.\ref{N1}(4), $F_1$ and $F_2$ represent $\chi^\pm$. $S$ denotes CP-even(CP-odd) scalar neutrino $\tilde{\nu}^{R(I)}$. $m_{F_1}$ and $m_{F_2}$ are the chargino masses. The concrete forms of the couplings of chargino-scalar neutrino-lepton and chargino-Z-chargino are collected in Eq.(\ref{A4}) of the Appendix. For Fig.\ref{N1}(5), $F_1$ and $F_2$ represent neutralinos. $S$ denotes scalar lepton. $m_{F_1}$ and $m_{F_2}$ are the neutralino masses. The corresponding couplings are in Eq.(\ref{A5}) of the Appendix. For Fig.\ref{N1}(6), $F_1$ and $F_2$ represent neutrinos. $S$ denotes charged Higgs. $m_{F_1}$ and $m_{F_2}$ denote the neutrino masses. We show the couplings in Eq.(\ref{A6}).

The specific form of the one-loop functions $G_i(x_1,x_2,x_3)(i=1...3)$ are
\begin{eqnarray}
&&G_1(x_1,x_2,x_3)\!=\!\frac{1}{16\pi^2}[\frac{x_1\ln{x_1}}{(x_1\!-\!x_2)(x_1\!-\!x_3)}+\frac{x_2\ln{x_2}}{(x_2\!-\!x_1)(x_2\!-\!x_3)}+\frac{x_3\ln{x_3}}{(x_3\!-\!x_1)(x_3\!-\!x_2)}],\nonumber\\
&&G_2(x_1,x_2,x_3)\!=\!\frac{1}{16\pi^2}[\frac{x_1^2\ln{x_1}}{(x_1\!-\!x_2)(x_1\!-\!x_3)}+\frac{x_2^2\ln{x_2}}{(x_2\!-\!x_1)(x_2\!-\!x_3)}+\frac{x_3^2\ln{x_3}}{(x_3\!-\!x_1)(x_3\!-\!x_2)}].
\end{eqnarray}

 The contributions obtained from Fig.\ref{N1}(7) and Fig.\ref{N1}(8) are expressed by $F_{L,R}(W)=F^{\alpha}_{L,R}(W)(\alpha=1,2)$. The specific forms are as follows:
\begin{eqnarray}
&&F^{(1,2)}_L(W)=i\Big[3H^{W_2F\overline{l}_i}_LH^{Z{W_1}W^{\ast}_2}H^{W^{\ast}_1{l_j}\overline{F}}_LG_2(x_F,x_{W_1},x_{W_2})\nonumber\\
&&\hspace{2.2cm}-H^{WF_2\overline{l}_i}_LH^{Z{F_1}\overline{F}_2}_LH^{\overline{F}_1{l_j}W^{\ast}}_LG_2(x_W,x_{F_1},x_{F_2})\Big],\nonumber\\
&&F^{(1,2)}_R(W)=0.
\end{eqnarray}

Here, $F(F_1, F_2)$ represents neutrino. The needed couplings of W-Z-W, lepton-neutrino-W and neutrino-Z-neutrino are collected in the Eqs.(\ref{A7}) and (\ref{A8}) of the Appendix.

 The contributions obtained from Fig.\ref {N2} are expressed by $F_{L,R}(B)=F^{\alpha}_{L,R}(B)(\alpha=1...4)$. The specific forms are as follows:
\begin{eqnarray}
&&F^{(1,2)}_L(B)=\frac{H^{Zl_i\overline{l}_i}_L}{m^2_{l_j}-m^2_{l_i}}\Big\{I_1(x_F,x_S)+\frac{m^2_{l_j}}{m^2_{W}}[I_2(x_F,x_S)-I_3(x_F,x_S)]\nonumber\\
&&\hspace{2.2cm}(m_{l_j}m_{F}H^{SF\overline{l}_i}_RH^{S^{\ast}{l_j}\overline{F}}_R+m_{l_i}m_{F}H^{SF\overline{l}_i}_LH^{S^{\ast}{l_j}\overline{F}}_L)\nonumber\\
&&\hspace{2.2cm}-\frac{1}{2}G_3(x_F,x_S)(m^2_{l_j}H^{SF\overline{l}_i}_RH^{S^{\ast}{l_j}\overline{F}}_L+m_{l_i}m_{l_j}H^{SF\overline{l}_i}_LH^{S^{\ast}{l_j}\overline{F}}_R)\Big\},\nonumber\\
&&F^{(3,4)}_L(B)=\frac{H^{Zl_j\overline{l}_j}_L}{m^2_{l_i}-m^2_{l_j}}\Big\{I_1(x_F,x_S)+\frac{m^2_{l_i}}{m^2_{W}}[I_2(x_F,x_S)-I_3(x_F,x_S)]\nonumber\\
&&\hspace{2.2cm}(m_{l_i}m_{F}H^{SF\overline{l}_i}_RH^{S^{\ast}{l_j}\overline{F}}_R+m_{l_j}m_{F}H^{SF\overline{l}_i}_LH^{S^{\ast}{l_j}\overline{F}}_L)\nonumber\\
&&\hspace{2.2cm}-\frac{1}{2}G_3(x_F,x_S)(m^2_{l_i}H^{SF\overline{l}_i}_LH^{S^{\ast}{l_j}\overline{F}}_R+m_{l_i}m_{l_j}H^{SF\overline{l}_i}_RH^{S^{\ast}{l_j}\overline{F}}_L)\Big\},\nonumber\\
&&F^{\alpha}_R(B)=F^{\alpha}_L(B)|_{L\leftrightarrow{R}},~~~\alpha=1...4.
\end{eqnarray}

$H^{Zl_i\overline{l}_i}_L=H^{Zl_j\overline{l}_j}_L=\frac{i}{2}(-g_1\cos\theta_W^\prime\sin\theta_W+g_2\cos\theta_W\cos\theta_W^\prime+g_{YX}\sin\theta_W^\prime)$
represents left-handed coupling of lepton-Z-lepton. For the Fig.\ref{N2}(1), $F$ and $S$  denote chargino and CP-even(CP-odd) scalar neutrino. $m_F$ represents the mass of chargino.
For the Fig.\ref{N2}(2), $F$ and $S$ denote neutralino and scalar lepton. $m_F$ is the mass of neutralino. The needed couplings for Fig.\ref{N2}(1) and Fig.\ref{N2}(2) can be found from the couplings for the Fig.\ref{N1}(1) and Fig.\ref{N1}(2). The conditions of Fig.\ref{N2}(3) and Fig.\ref{N2}(4) are similar as those of Fig.\ref{N2}(1) and Fig.\ref{N2}(2).

Here
\begin{eqnarray}
&&I_1(x_1,x_2)=\frac{1}{16\pi^2}\Big[\frac{1+\log{x_2}}{x_2-x_1}+\frac{x_1\log{x_1}-x_2\log{x_2}}{(x_2-x_1)^2}\Big],\nonumber\\
&&I_2(x_1,x_2)=\frac{1}{16\pi^2}\Big[-\frac{1+\log{x_2}}{x_2-x_1}-\frac{x_1\log{x_1}-x_2\log{x_2}}{(x_2-x_1)^2}\Big],\nonumber\\
&&G_3(x_1,x_2)=-\frac{1}{16\pi^2}\Big[\frac{x^2_2\log{x_2}-x^2_1\log{x_1}}{(x_2-x_1)^2}+\frac{x_2+2x_2\log{x_2}}{(x_1-x_2)}-\frac{1}{2}\Big].
\end{eqnarray}

 Then, the branching ratios of $Z\rightarrow{{l_i}^{\pm}{l_j}^{\mp}}$ are defined as
\begin{eqnarray}
&&Br(Z\rightarrow{{l_i}^{\pm}{l_j}^{\mp}})=\frac{1}{12\pi}\frac{m_Z}{\Gamma_Z}(|F^Z_L|^2+|F^Z_R|^2)\nonumber\\
&&\hspace{3cm}=\frac{1}{12\pi}\frac{m_Z}{\Gamma_Z}(|F_L(A)+F_L(W)+F_L(B)|^2+|F_R(A)+F_R(B)|^2),
\end{eqnarray}
here $\Gamma_Z$ represents the total decay width of the Z-boson, $\Gamma_Z\simeq2.4952$ GeV\cite{IN4}.

\section{Higgs boson decays $h\rightarrow{{l_i}^{\pm}{l_j}^{\mp}}$}
\begin{figure}[h]
\setlength{\unitlength}{5.0mm}
\centering
\includegraphics[width=5.0in]{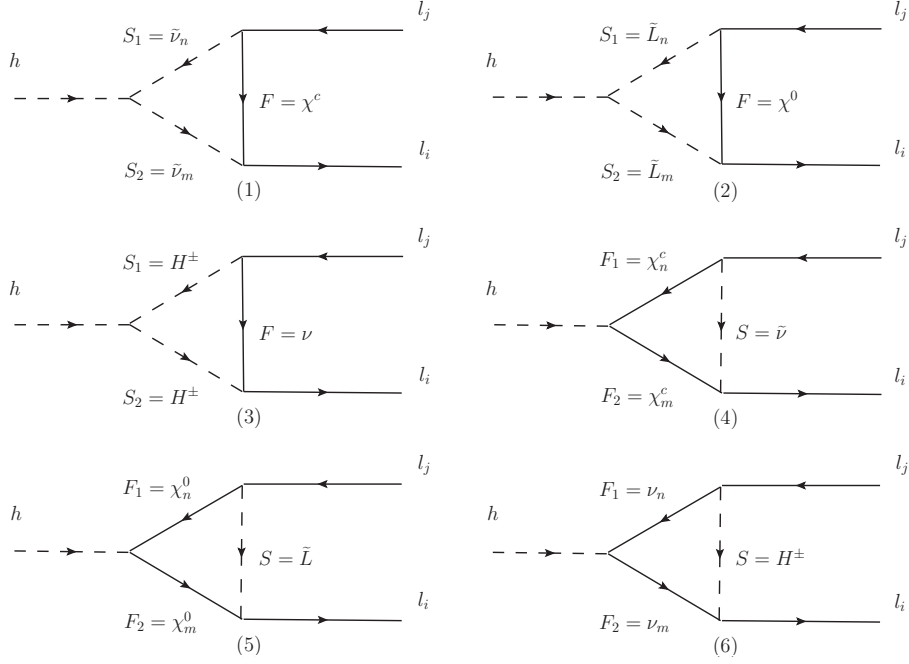}
\caption{ Feynman diagrams for the processes $h\rightarrow{{l_i}^{\pm}{l_j}^{\mp}}$ in the $U(1)_X$SSM, which denote the contributions of vertex diagrams for $h\rightarrow{{l_i}^{\pm}{l_j}^{\mp}}$ from loops.}\label{N3}
\end{figure}

\begin{figure}[h]
\setlength{\unitlength}{5.0mm}
\centering
\includegraphics[width=5.0in]{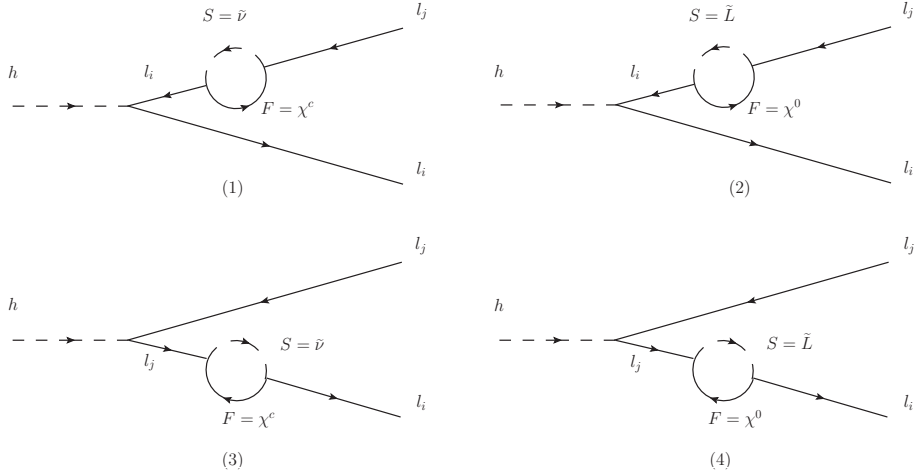}
\caption{ Feynman diagrams for the processes $h\rightarrow{{l_i}^{\pm}{l_j}^{\mp}}$ in the $U(1)_X$SSM, which denote self-energy diagrams contributing to  $h\rightarrow{{l_i}^{\pm}{l_j}^{\mp}}$ from loops.}\label{N4}
\end{figure}

 In this section, we analyze the LFV processes $h\rightarrow{{l_i}^{\pm}{l_j}^{\mp}}$. The corresponding Feynman diagrams can be depicted by Fig.\ref {N3} and Fig.\ref {N4}.

 The corresponding effective amplitude can be written as
\begin{eqnarray}
\mathcal{M}=\overline{l}_i(F_LP_L+F_RP_R){l_j}h,
\end{eqnarray}
with
\begin{eqnarray}
F^h_{L,R}=F_{L,R}(C)+F_{L,R}(D),
\end{eqnarray}

The contribution obtained by Fig.\ref {N3} is expressed by $F_{L,R}(C)=F^{\alpha}_{L,R}(C)(\alpha=1...6)$. The specific forms are as follows:
\begin{eqnarray}
&&F^{(1,2,3)}_L(C)=\frac{m_F}{m^2_{N_p}}H^{h{S_1}S^{\ast}_2}H^{S_2F\overline{l}_i}_LH^{S^{\ast}_1{l_j}\overline{F}}_LG_1(x_F,x_{S_1},x_{S_2}),\nonumber\\
&&F^{(4,5,6)}_L(C)=\frac{m_{F_1}m_{F_2}}{m^2_{N_p}}H^{SF_2\overline{l}_i}_LH^{h{F_1}\overline{F}_2}_LH^{S^{\ast}{l_j}\overline{F}_1}_LG_1(x_S,x_{F_1},x_{F_2})\nonumber\\
&&\hspace{2.4cm}+H^{SF_2\overline{l}_i}_LH^{h{F_1}\overline{F}_2}_RH^{S^{\ast}{l_j}\overline{F}_1}_LG_2(x_S,x_{F_1},x_{F_2}),\nonumber\\
&&F^{\alpha}_R(C)=F^{\alpha}_L(C)|_{L\leftrightarrow{R}},~~~\alpha=1...6.
\end{eqnarray}

The Figs.\ref{N3}(1), \ref{N3}(2) \dots \ref{N3}(6) are similar as the Figs.\ref{N1}(1), \ref{N1}(2) \dots \ref{N1}(6) with Z replaced by h. So we just show the couplings relating with h. For Fig.\ref{N3}(1) $H^{h{S_1}S^{\ast}_2}\rightarrow H^{h\tilde{\nu}^R\tilde{\nu}^{R*}}(H^{h\tilde{\nu}^I\tilde{\nu}^{I*}})$. The concrete form of $H^{h\tilde{\nu}^R\tilde{\nu}^{R*}}$ can be found in the appendix (Eq.(A10)) of Ref.\cite{UU3}. $H^{h\tilde{\nu}^I\tilde{\nu}^{I*}}$ is very similar as $H^{h\tilde{\nu}^R\tilde{\nu}^{R*}}$ with the replacement $Z^R\rightarrow Z^I$.

For Fig.\ref{N3}(2), $S_1$ and $S_2$ denote scalar lepton.
Then  $H^{h{S_1}S^{\ast}_2}\rightarrow H^{h\tilde{L}\tilde{L}^*}$, which reads as
\begin{eqnarray}
&&H^{h\tilde{L}_n\tilde{L}^*_m}=\frac{i}{4}\Big\{\sum_{a=1}^3Z_{m,a}^{E,*}Z_{n,a}^E
\Big((g_2^2-g_{YX}g_X-g_1^2-g_{YX}^2)(v_dZ_{b1}^H-v_uZ_{b2}^H)+g_{YX}g_X(v_{\overline{\eta}}Z_{b4}^H\nonumber\\
&&\hspace{1.6cm}-v_{\eta}Z_{b3}^H)\Big)+\sum_{a=1}^3Z_{m,3+a}^{E,*}Z_{n,3+a}^E\Big((2g_1^2+2g_{YX}^2+3g_{YX}g_X+g_X^2)(v_dZ_{b1}^H-v_uZ_{b2}^H)\nonumber\\
&&\hspace{1.6cm}+2(g_{YX}g_X+g_X^2)(-v_{\overline{\eta}}Z_{b4}^H+v_{\eta}Z_{b3}^H)\Big)+\Big(\sum_{a=1}^3Z_{m,a}^{E,*}Z_{n,3+a}^E+\sum_{a=1}^3Z_{m,3+a}^{E,*}Z_{n,a}^E\Big)\nonumber\\
&&\hspace{1.6cm}\times\Big[-2\sqrt{2}T_{e,a}Z_{b1}^H+Y_{e,a}\Big(2(v_S\lambda_H+\sqrt{2}{\mu})Z_{b2}^H+2v_u\lambda_HZ_{b5}^H\Big)\Big]\Big\}.
\end{eqnarray}

For Fig.\ref{N3}(3), the scalar particle is charged Higgs and
\begin{eqnarray}
&&H^{h{S_1}S^{\ast}_2}\rightarrow H^{hH^\pm_m H^{\pm*}_n}\nonumber\\
&&=\frac{i}{4}\Big\{(-Z_{b2}^HZ_{m2}^+-Z_{b1}^HZ_{m1}^+)\Big([(g_{YX}+g_X)^2+g_1^2+g_2^2](v_uZ_{n2}^++v_dZ_{n1}^+)
+(g_2^2-2\lambda_H^2)\nonumber\\
&&\times(v_dZ_{n1}^+-v_uZ_{n2}^+)\Big)
+(Z_{b2}^HZ_{m1}^++Z_{b1}^HZ_{m2}^+)\Big([(g_{YX}+g_X)^2-2g_2^2+g_1^2+2\lambda_H^2](v_uZ_{n1}^+\nonumber\\
&&+v_dZ_{n2}^+)\Big)-2Z_{b4}^H(Z_{m2}^+-Z_{m1}^+)\Big((g_{YX}g_X+g_X^2)v_{\overline{\eta}}(Z_{n2}^++Z_{n1}^+)
+\lambda_cv_{\eta}\lambda_H(Z_{n1}^+-Z_{n2}^+)\Big)\nonumber\\&&+Z_{b3}^H(Z_{m2}^++Z_{m1}^+)
\Big((g_{YX}g_X+g_X^2)(v_{\eta}Z_{n1}^+-v_{\eta}Z_{n2}^+)+\lambda_cv_{\overline{\eta}}\lambda_H^*(Z_{n1}^++Z_{n2}^+)\Big)
\nonumber\\
&&+Z_{b5}^H(Z_{m2}^++Z_{m1}^+)(Z_{n2}^++Z_{n1}^+)\Big(\sqrt{2}T_{{\lambda},H}+2\lambda_H(\kappa v_s+\sqrt{2}M_S
+\sqrt{2}{\mu}+\lambda_Hv_S)\Big)\Big\}.
\end{eqnarray}

For Fig.\ref{N3}(4), $F_1(F_2)$ denotes chargino, while $m_{F_1}(m_{F_2})$ represents chargino mass.
The corresponding couplings are
\begin{eqnarray}
&& H^{h{F_1}\overline{F}_2}_L\rightarrow H^{h\chi^\pm_n\overline{\chi}^\pm_m}_L=-\frac{i}{\sqrt{2}}\Big(g_2U_{m1}^*V_{n2}^*Z_{b2}^H
 +U_{m2}^*(g_2V_{n1}^*Z_{b1}^H+\lambda_HV_{n2}^*Z_{b5}^H)\Big),\nonumber\\&&
 H^{h{F_1}\overline{F}_2}_R\rightarrow H^{h\chi^\pm_n\overline{\chi}^\pm_m}_R=-\frac{i}{\sqrt{2}}\Big(g_2U_{n1}V_{m2}Z_{b2}^H+U_{n2}(g_2V_{m1}Z_{b1}^H+\lambda_H^*V_{m2}Z_{b5}^H)\Big).
\end{eqnarray}

For Fig.\ref{N3}(5), $F_1(F_2)$ and $m_{F_1}(m_{F_2})$ represent neutralino and neutralino mass respectively. The Higgs coupling with neutralino $H^{h\chi^0\overline{\chi}^0}$ is shown as Eq.(A5) in the appendix of our previous work\cite{UU5}. For Fig.\ref{N3}(6), $F_1$ and $F_2$ are neutrinos. The terms proportional to tiny neutrino mass $(m_{F_1}, m_{F_2})$ are not of interest. We do not show the higgs-neutrino-neutrino couplings, because the corrections from Fig.\ref{N3}(6) are very small.

The contributions obtained from Fig.\ref {N4} are expressed by $F_{L,R}(D)=F^{\alpha}_{L,R}(D)(\alpha=1...4)$. The specific forms are as follows:
\begin{eqnarray}
&&F^{(1,2)}_L(D)=\frac{H^{hl_i\overline{l}_i}_L}{m^2_{l_j}-m^2_{l_i}}\Big\{I_1(x_F,x_S)+\frac{m^2_{l_j}}{m^2_{W}}[I_2(x_F,x_S)-I_3(x_F,x_S)]\nonumber\\
&&\hspace{2.2cm}(m_{l_j}m_{F}H^{SF\overline{l}_i}_RH^{S^{\ast}{l_j}\overline{F}}_R+m_{l_i}m_{F}H^{SF\overline{l}_i}_LH^{S^{\ast}{l_j}\overline{F}}_L)\nonumber\\
&&\hspace{2.2cm}-\frac{1}{2}G_3(x_F,x_S)(m^2_{l_j}H^{SF\overline{l}_i}_RH^{S^{\ast}{l_j}\overline{F}}_L+m_{l_i}m_{l_j}H^{SF\overline{l}_i}_LH^{S^{\ast}{l_j}\overline{F}}_R)\Big\},\nonumber\\
&&F^{(3,4)}_L(D)=\frac{H^{hl_j\overline{l}_j}_L}{m^2_{l_i}-m^2_{l_j}}\Big\{I_1(x_F,x_S)+\frac{m^2_{l_i}}{m^2_{W}}[I_2(x_F,x_S)-I_3(x_F,x_S)]\nonumber\\
&&\hspace{2.2cm}(m_{l_i}m_{F}H^{SF\overline{l}_i}_RH^{S^{\ast}{l_j}\overline{F}}_R+m_{l_j}m_{F}H^{SF\overline{l}_i}_LH^{S^{\ast}{l_j}\overline{F}}_L)\nonumber\\
&&\hspace{2.2cm}-\frac{1}{2}G_3(x_F,x_S)(m^2_{l_i}H^{SF\overline{l}_i}_LH^{S^{\ast}{l_j}\overline{F}}_R+m_{l_i}m_{l_j}H^{SF\overline{l}_i}_RH^{S^{\ast}{l_j}\overline{F}}_L)\Big\},\nonumber\\
&&F^{\alpha}_R(D)=F^{\alpha}_L(D)|_{L\leftrightarrow{R}},~~~\alpha=1...4.
\end{eqnarray}

The lepton-h-lepton coupling is denoted by $H_L^{hl_i\overline{l}_i}=-\frac{i}{\sqrt{2}}Y_{e,i}Z_{b1}^H$. In the Fig.\ref{N4}, the other couplings and $m_F$ are same as the corresponding terms in the Fig.\ref{N2}.

 Then, the branching ratio of $h\rightarrow{{l_i}^{\pm}{l_j}^{\mp}}$ is defined as
\begin{eqnarray}
Br(h\rightarrow{{l_i}^{\pm}{l_j}^{\mp}})=\frac{1}{16\pi}\frac{m_h}{\Gamma_h}(|F^h_L|^2+|F^h_R|^2),
\end{eqnarray}
here $\Gamma_h\simeq\Gamma^{SM}_h\simeq4.1\times10^{-3}$ GeV\cite{Z184}. $\Gamma_h$ represents the total decay width of the Higgs boson in the $U(1)_X$SSM. $\Gamma^{SM}_h$
represents the predicted value of the 125 GeV Higgs boson total decay width in the SM. In the following numerical section, we choose the supersymmetric particles in the $U(1)_X$SSM that are heavy and whose contributions to the decay width of the 125 GeV Higgs boson is
weak. Hence, we choose $\Gamma_h$ which is approximately equal to $\Gamma^{SM}_h$.

\section{Numerical analysis}
 In this section, we study the numerical results and consider the experiments constraints from the lightest CP-even Higgs mass $m_{h^0}$=125.1 GeV\cite{IN2,IN3,xin1,ZPG1,ZPG2,TanBP}.
 In order to obtain reasonable numerical results, we need to study some sensitive parameters. We need to consider the effect of $l_j\rightarrow{l_i\gamma}$ on LFV. The limitation of
$\mu\rightarrow{e\gamma}$ is the strongest, and other restrictions can be achieved if the limit of $\mu\rightarrow{e\gamma}$ is satisfied\cite{T1}. Then, to show the numerical results clearly, we will discuss the processes of $Z\rightarrow e\mu$,~$Z\rightarrow e\tau$,~$Z\rightarrow \mu\tau$ and $h\rightarrow e\mu$,~$h\rightarrow e\tau$,~$h\rightarrow \mu\tau$~in six subsections. We draw the relation diagrams and scatter diagrams with different parameters. After analyzing these graphs and the experimental limits of the branching ratios, reasonable parameter spaces are found to explain LFV.

 According to the latest LHC data\cite{w1,w2,w3,w4,w5,w6}, we take for the scalar lepton mass greater than $700~{\rm GeV}$, the chargino mass greater than $1100~{\rm GeV}$, and the scalar quark mass greater than $1500~{\rm GeV}$. $M_{Z^{\prime}}> 5.1$ TeV is the latest experimental constraint on the mass of the added
 heavy vector boson $Z^\prime$\cite{xin1}. The upper bound of the ratio of $Z^\prime$ mass to its gauge coupling $M_{Z^\prime}/g_X\geq6$ TeV under 99\% C.L.\cite{ZPG1,ZPG2} is given in the references. Taking into account the constraint from LHC data, $\tan \beta_\eta<1.5$\cite{TanBP}. Combined with the above experimental requirements, we get abundant data, and process the data to get interesting one-dimensional graphs and multidimensional scatter plots. Considering the above constraints in the front paragraph, we use the following parameters
\begin{eqnarray}
&&g_X=0.3,~g_{YX}=0.1,~\lambda_H = 0.1,~\lambda_C = -0.2,~\sqrt{v_{\eta}^2+v_{\overline{\eta}}^2}=17~{\rm TeV},\nonumber\\
&&{\mu}=M_{BL}=T_{\lambda_H}=T_{\lambda_C}=T_{\kappa} = 1~{\rm TeV},~M_{BB^\prime}=0.4~{\rm TeV},~\kappa=0.1,\nonumber\\
&&l_W = B_{\mu} =B_S=0.1~{\rm TeV}^2,~T_{Xii}= -1~{\rm TeV},~Y_{Xii}= 1,~(i=1,2,3).
\end{eqnarray}

 To simplify the numerical research, we use the relations for the parameters and they vary in the following numerical analysis
\begin{eqnarray}
&&\hspace{1.3cm}M^2_{\tilde{L}ij}=M^2_{\tilde{L}ji},~~M^2_{\tilde{E}ij}=M^2_{\tilde{E}ji},\nonumber\\
&&M^2_{\tilde{\nu}ij}=M^2_{\tilde{\nu}ji},~~T_{eij}=T_{eji},~~T_{{\nu}ij}=T_{{\nu}ji}(i\ne j).
\end{eqnarray}

Generally, the non-diagonal elements of the parameters are defined as zero unless we note otherwise.

\subsection{$Z\rightarrow e{\mu}$}

\begin{figure}[h]
\setlength{\unitlength}{5mm}
\centering
\includegraphics[width=3.0in]{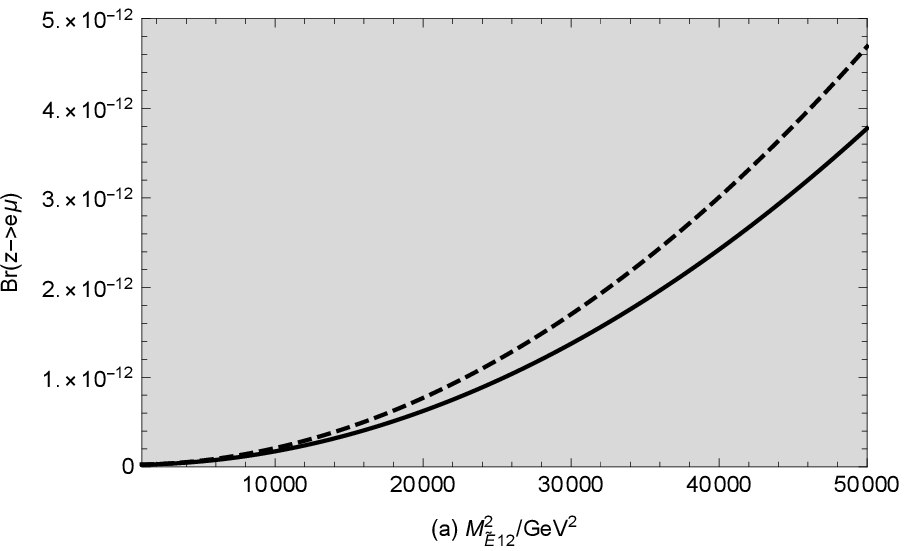}
\vspace{0.2cm}
\setlength{\unitlength}{5mm}
\centering
\includegraphics[width=3.0in]{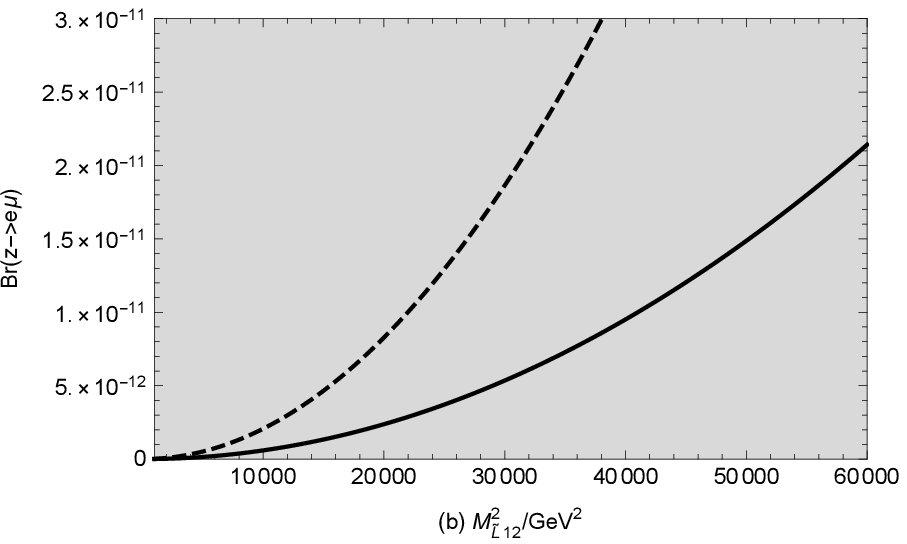}
\vspace{0.2cm}
\setlength{\unitlength}{5mm}
\centering
\includegraphics[width=3.0in]{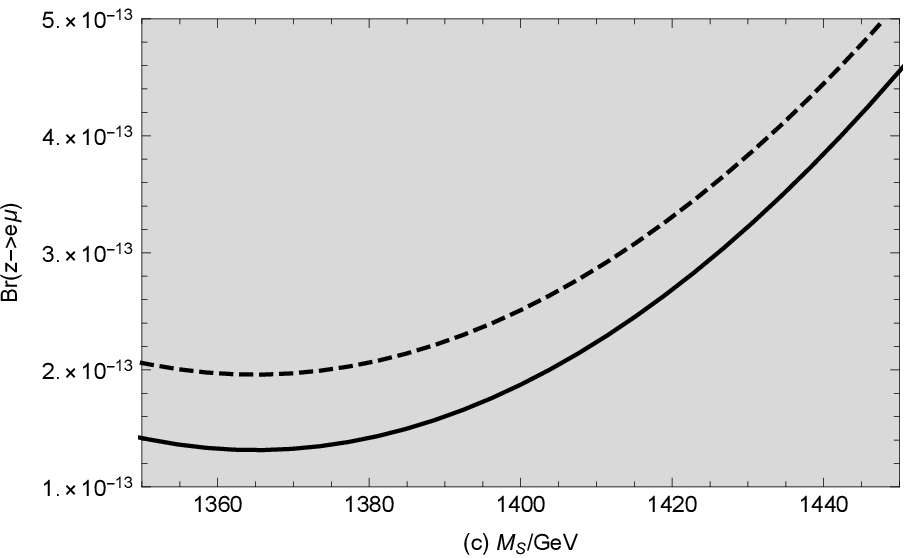}
\caption{$Br(Z\rightarrow e{\mu})$ schematic diagrams affected by different parameters. The gray area is reasonable value range where $Br(Z\rightarrow e{\mu})$ satisfies the upper limit. The dashed and solid lines in Fig.\ref {N5}(a)(b) correspond to $M_S =1.5~{\rm TeV}$ and $M_S =1.2~{\rm TeV}$. The dashed and solid lines in Fig.\ref {N5}(c) correspond to $M^2_{\tilde{L}12}=6\times10^{3}~{\rm GeV}^2$ and $M^2_{\tilde{L}12}=5\times10^{3}~{\rm GeV}^2$.}\label{N5}
\end{figure}

With the parameters $v_S=4.3~{\rm TeV}$,~$M_1 =1.2~{\rm TeV}$,~$M^2_{\tilde{L}ii}=3~{\rm TeV}^2$,~$T_{eii}=0.5~{\rm TeV}$,~$T_{{\nu}ii}=1~{\rm TeV}$,~$M^2_{{\tilde{\nu}}ii}=0.3~{\rm TeV}^2$,~$M^2_{\tilde{E}ii}=0.8~{\rm TeV}^2$,~(i=1,2,3),
we paint $Br(Z\rightarrow e{\mu})$ schematic diagrams affected by different parameters in the Fig.\ref {N5}. The gray area is the experimental limit that this process satisfies.

 In the Fig.\ref {N5}(a), we plot $Br(Z\rightarrow e{\mu})$ versus $M^2_{\tilde{E}12}$, in which the dashed curve corresponds to $M_S =1.5~{\rm TeV}$ and the solid line corresponds to
$M_S =1.2~{\rm TeV}$. We can clearly see that the two lines increase with the increasing $M^2_{\tilde{E}12}$ in the range of $10^{3}~{\rm GeV}^2-5\times10^{4}~{\rm GeV}^2$. The dashed curve is larger than the solid curve. The solid line and the dashed line are located in the gray area. In the Fig.\ref {N5}(b), we plot $Br(Z\rightarrow e{\mu})$ versus $M^2_{\tilde{L}12}$, which the dashed curve corresponds to $M_S =1.5~{\rm TeV}$ and the solid line corresponds to
$M_S =1.2~{\rm TeV}$. We can clearly see that the two lines increase with the increasing $M^2_{\tilde{L}12}$ in the range of $10^{3}~{\rm GeV}^2-6\times10^{4}~{\rm GeV}^2$. The dashed curve is also larger than the solid curve. Both the solid line and the dashed line are located in the gray area. In the Fig.\ref {N5}(c), we plot $Br(Z\rightarrow e{\mu})$ versus $M_S$, in which the dashed curve corresponds to $M^2_{\tilde{L}12}=6\times10^{3}~{\rm GeV}^2$ and the solid line corresponds to
$M^2_{\tilde{L}12}=5\times10^{3}~{\rm GeV}^2$. We can clearly see that the two lines increase with the increasing $M_S$ in the range of $1350~{\rm GeV}-1450~{\rm GeV}$. The dashed curve is larger than the solid curve. The solid line and the dashed line are located in the gray area. For the other fixed parameters, it is based on our previous works especially for the LFV processes $l_j\rightarrow{l_i\gamma}$\cite{T1,T10} in $U(1)_XSSM$.
The constraint from $\mu\rightarrow e\gamma$ is strict. The other restrictions are relatively loose and easy to be satisfied.

\begin{figure}[h]
\setlength{\unitlength}{5mm}
\centering
\includegraphics[width=3.0in]{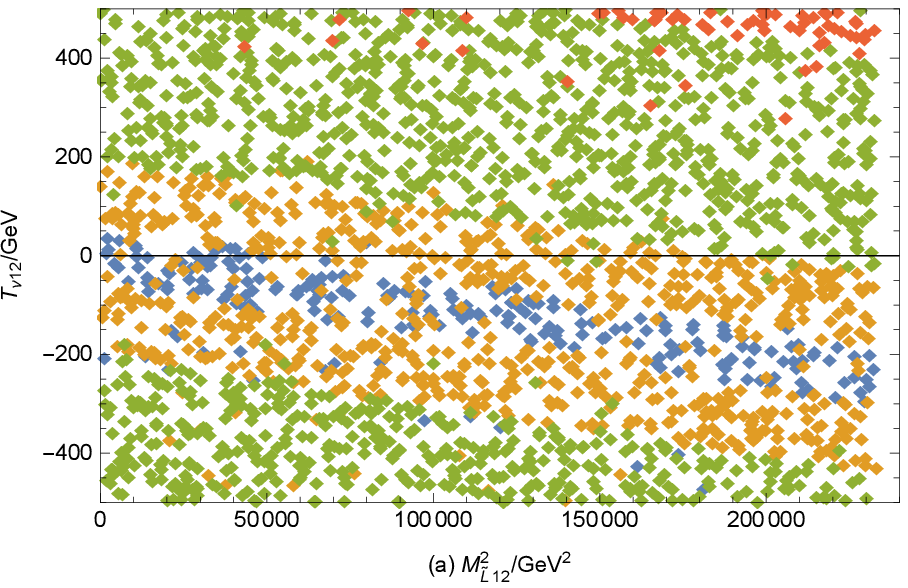}
\vspace{0.2cm}
\setlength{\unitlength}{5mm}
\centering
\includegraphics[width=3.0in]{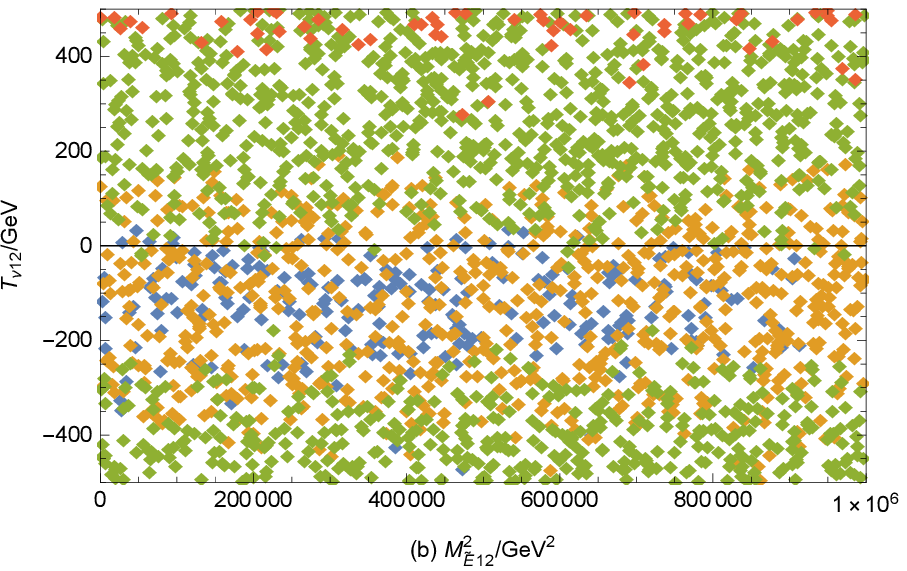}
\vspace{0.2cm}
\setlength{\unitlength}{5mm}
\centering
\includegraphics[width=3.0in]{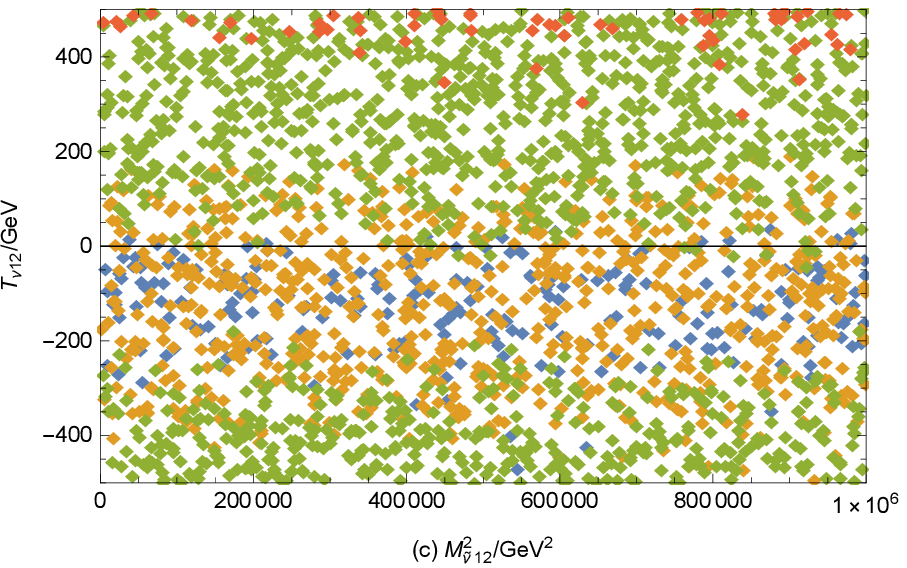}
\caption{Under the premise of current limit on lepton flavor violating decay $Z\rightarrow e{\mu}$. Reasonable parameter space is selected to scatter points, with the notation blue (0$<Br(Z\rightarrow e{\mu})<5\times10^{-14}$), yellow ($5\times10^{-14}\leq Br(Z\rightarrow e{\mu})<5\times10^{-13}$), green ($5\times10^{-13}$ $\leq Br(Z\rightarrow e{\mu})<5\times10^{-12}$) and red ($5\times10^{-12}$ $\leq Br(Z\rightarrow e{\mu})<7.5\times10^{-7}$).}
{\label {N6}}
\end{figure}
In summary, $M^2_{\tilde{E}12}$ and $M^2_{\tilde{L}12}$ are the flavor mixing parameters appearing in the mass matrixes of the slepton, CP-even sneutrino and CP-odd sneutrino.
The mass for the super partner of the Higgs singlet $S$ is denoted by $M_S$ included  in the mass matrixes of Higgs and neutralino.
So the contributions can be influenced to some extent  by the parameters $M^2_{\tilde{E}12}$, $M^2_{\tilde{L}12}$ and $M_S$. $Br(Z\rightarrow e{\mu})$ increases as the parameters $M^2_{\tilde{E}12}$, $M^2_{\tilde{L}12}$ and $M_S$ increase. In the Fig.\ref {N5}, the dashed line has a higher slope than the solid line, and they vary in the region $10^{-13}- 10^{-11}$ much smaller than its current limit. All in all, $M^2_{\tilde{E}12}$, $M^2_{\tilde{L}12}$ and $M_S$ are sensitive parameters
that have obvious effects on $Br(Z\rightarrow e{\mu})$.
\begin{table*}
\caption{Scanning parameters for Fig.{\ref {N6}} and Fig.{\ref {N11}} with i=1,2,3}\label{V}
\begin{tabular}{|c|c|c|}
\hline
Parameters&Min&Max\\
\hline
$\hspace{1.5cm}M^2_{\tilde{L}12}/\rm GeV^2\hspace{1.5cm}$ &$\hspace{1.5cm}0\hspace{1.5cm}$& $\hspace{1.5cm}10^6\hspace{1.5cm}$\\
\hline
$\hspace{1.5cm}M^2_{\tilde{E}12}/\rm GeV^2\hspace{1.5cm}$ &$\hspace{1.5cm}0\hspace{1.5cm}$& $\hspace{1.5cm}10^6\hspace{1.5cm}$\\
\hline
$\hspace{1.5cm}M^2_{\tilde{\nu}12}/\rm GeV^2\hspace{1.5cm}$ &$\hspace{1.5cm}0\hspace{1.5cm}$ &$\hspace{1.5cm}10^6\hspace{1.5cm}$\\
\hline
$T_{e12}$/GeV & $\hspace{1.5cm}-300\hspace{1.5cm}$ &$\hspace{1.5cm}300\hspace{1.5cm}$\\
\hline
$T_{{\nu}12}$/GeV &$\hspace{1.5cm}-500\hspace{1.5cm}$ &$\hspace{1.5cm}500\hspace{1.5cm}$\\
\hline
$\hspace{1.5cm}M^2_{\tilde{L}ii}/\rm GeV^2\hspace{1.5cm}$ &$\hspace{1.5cm}2\times10^{5}\hspace{1.5cm}$& $\hspace{1.5cm}10^9\hspace{1.5cm}$\\
\hline
$\hspace{1.5cm}M^2_{\tilde{E}ii}/\rm GeV^2\hspace{1.5cm}$ &$\hspace{1.5cm}2\times10^{5}\hspace{1.5cm}$& $\hspace{1.5cm}10^9\hspace{1.5cm}$\\
\hline
$\hspace{1.5cm}M^2_{\tilde{\nu}ii}/\rm GeV^2\hspace{1.5cm}$ &$\hspace{1.5cm}1\times10^{5}\hspace{1.5cm}$ &$\hspace{1.5cm}10^9\hspace{1.5cm}$\\
\hline
$T_{eii}$/GeV & $\hspace{1.5cm}-3000\hspace{1.5cm}$ &$\hspace{1.5cm}3000\hspace{1.5cm}$\\
\hline
$T_{{\nu}ii}$/GeV &$\hspace{1.5cm}-3000\hspace{1.5cm}$ &$\hspace{1.5cm}3000\hspace{1.5cm}$\\
\hline
$\hspace{1.5cm}\tan\beta\hspace{1.5cm}$ &$\hspace{1.5cm}1\hspace{1.5cm}$ &$\hspace{1.5cm}50\hspace{1.5cm}$\\
\hline
$M_1$/GeV & $\hspace{1.5cm}200\hspace{1.5cm}$ &$\hspace{1.5cm}3000\hspace{1.5cm}$\\
\hline
$M_2$/GeV &$\hspace{1.5cm}600\hspace{1.5cm}$ &$\hspace{1.5cm}3000\hspace{1.5cm}$\\
\hline
\end{tabular}
\end{table*}

 Next, supposing the parameters with $M_S=1.2~{\rm TeV}$,  we randomly scan the parameters. All the parameters involved are expressed in tabular form. Fig.\ref {N6} is
 obtained from the parameters shown in the Table \ref {V}. We use blue (0$<Br(Z\rightarrow e{\mu})<5\times10^{-14}$), yellow ($5\times10^{-14}\leq Br(Z\rightarrow e{\mu})<5\times10^{-13}$), green ($5\times10^{-13}$ $\leq Br(Z\rightarrow e{\mu})<5\times10^{-12}$) and red ($5\times10^{-12}$ $\leq Br(Z\rightarrow e{\mu})<7.5\times10^{-7}$) to represent the results in different parameter spaces for the process of $Z\rightarrow e{\mu}$.

 The relationship between $M^2_{\tilde{L}12}$ and $T_{{\nu}12}$ is shown in Fig.\ref {N6}(a). We can see that the overall change trend of scattered points is obvious in Fig.\ref {N6}(a), where four types of points are concentrated in -500 GeV$<T_{{\nu}12}<500$ GeV. Blue parts are mainly in -300 GeV$<T_{{\nu}12}<50$ GeV, yellow parts are mainly  in -400 GeV$<T_{{\nu}12}<200$ GeV, green parts are mainly in -500 GeV$<T_{{\nu}12}<-200$ GeV and 0 GeV$<T_{\nu}<500$ GeV and red parts are mainly in 450 GeV$<T_{{\nu}12}<500$ GeV.

 The relationship between $M^2_{\tilde{E}12}$ and $T_{{\nu}12}$ is shown in Fig.\ref {N6}(b). The relationship between $M^2_{\tilde{\nu}12}$ and $T_{{\nu}12}$ is shown in Fig.\ref {N6}(c). In Fig.\ref {N6}(b) and Fig.\ref {N6}(c), we find that the variation trend of scattered points is weak, where blue parts are mainly in -200 GeV$<T_{{\nu}12}<0$ GeV, yellow parts are mainly in -380 GeV$<T_{{\nu}12}<100$ GeV, green parts are mainly in -500 GeV$<T_{{\nu}12}<-300$ GeV and 0 GeV$<T_{{\nu}12}<500$ GeV and red parts are mainly in 400 GeV$<T_{{\nu}12}<500$ GeV.

\subsection{$Z\rightarrow e{\tau}$}
\begin{figure}[h]
\setlength{\unitlength}{5mm}
\centering
\includegraphics[width=3.0in]{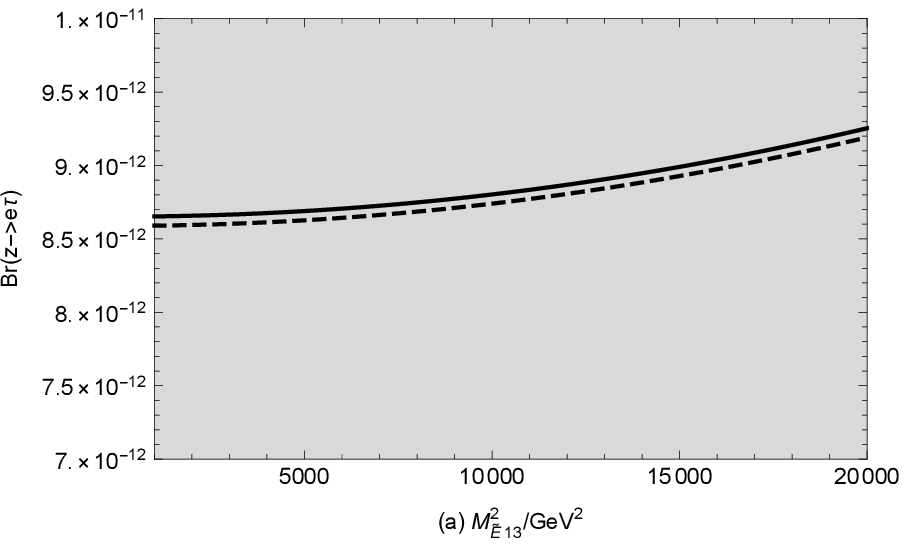}
\vspace{0.2cm}
\setlength{\unitlength}{5mm}
\centering
\includegraphics[width=3.0in]{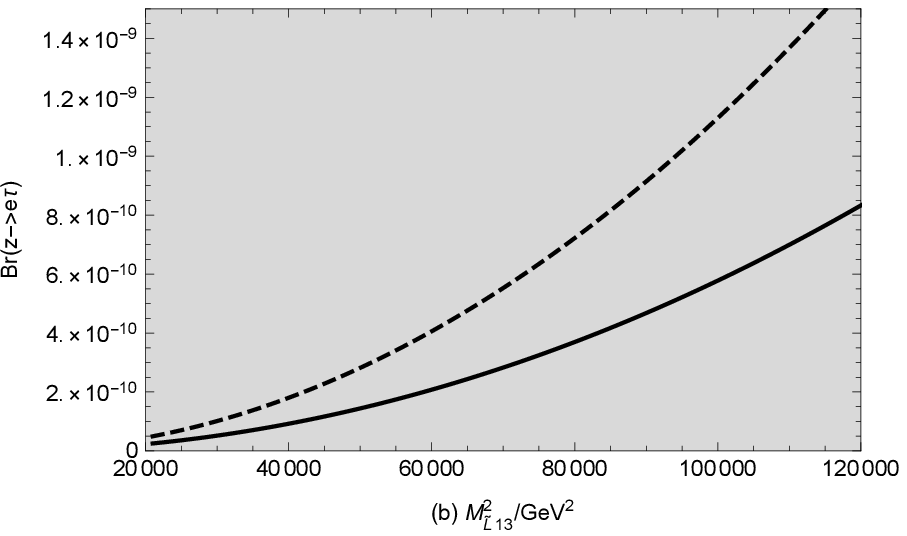}
\caption{$Br(Z\rightarrow e{\tau})$ schematic diagrams affected by different parameters. The gray area is reasonable value range, where $Br(Z\rightarrow e{\tau})$ satisfies the upper limit. The dashed and solid lines in Fig.\ref {N7}(a)(b) correspond to $M^2_{\tilde{L}ii}=2.5\times10^{6}~{\rm GeV^2}$ and $M^2_{\tilde{L}ii}=3\times10^{6}~{\rm GeV^2}$ with (i=1,2,3).}\label{N7}
\end{figure}

With the parameters $v_S=4.3~{\rm TeV}$,~$M_S =1.2~{\rm TeV}$,~$\tan{\beta}=20$,~$T_{eii}=2~{\rm TeV}$,~$T_{{\nu}ii}=3~{\rm TeV}$,~(i=1,2,3), we paint $Br(Z\rightarrow e{\tau})$ schematic diagrams affected by different parameters in the Fig.\ref {N7}. Identically, the gray area is the experimental limit that this process satisfies.

 We study the branching ratio of $Z\rightarrow e{\tau}$ versus $M^2_{\tilde{E}13}$ with $M^2_{\tilde{L}ii}=2.5\times10^{6}~{\rm GeV^2}(3\times10^{6}~{\rm GeV^2})$, (i=1,2,3).
In the Fig.\ref {N7}(a), the results are plotted by the dotted line and solid line respectively. The both line are almost overlap. The both lines increase with $M^2_{\tilde{E}13}$ increasing from $10^{3}$ $\rm GeV^2$ to $2\times10^{4}$ $\rm GeV^2$, which indicates that $M^2_{\tilde{E}13}$ is a sensitive parameter for the numerical results.
The solid line and the dashed line are located in the gray area. In the Fig.\ref {N7}(b), we plot $Z\rightarrow e{\tau}$ versus $M^2_{\tilde{L}13}$, in which the dashed curve corresponds to $M^2_{\tilde{L}ii}=2.5\times10^{6}~{\rm GeV^2}$ and the solid line corresponds to $M^2_{\tilde{L}ii}=3\times10^{6}~{\rm GeV^2}$, (i=1,2,3). The dashed curve is larger than the solid curve.
We can clearly see that the two lines increase with the increasing $M^2_{\tilde{L}13}$ in the range of $2\times10^{3}$ $\rm GeV^2$ to $1.2\times10^{4}$ $\rm GeV^2$. The solid line and the dashed line are located in the gray area. So, the contributions can be influenced by the parameters $M^2_{\tilde{E}13}$ and $M^2_{\tilde{L}13}$.

\begin{figure}[h]
\setlength{\unitlength}{5mm}
\centering
\includegraphics[width=3.0in]{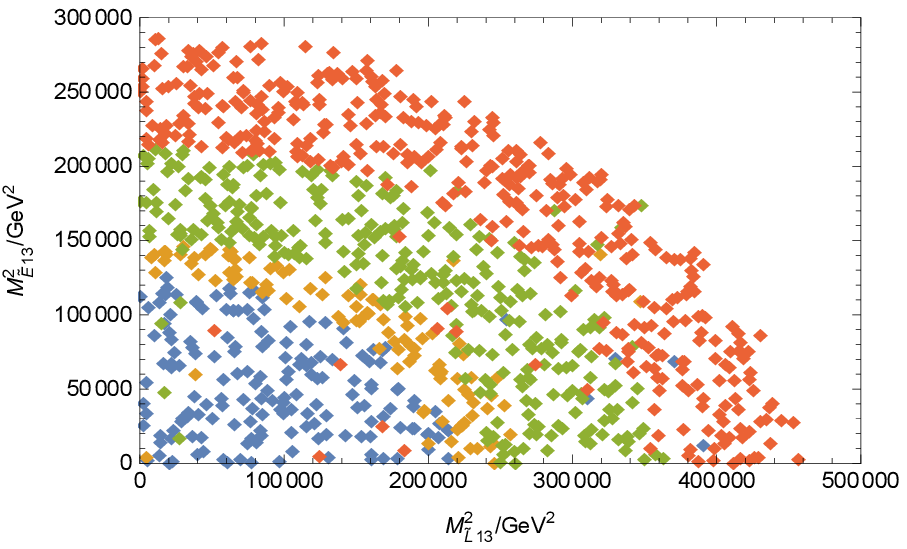}
\caption{Under the premise of current limit on lepton flavor violating decay $Z\rightarrow e{\tau}$. Reasonable parameter space is selected to scatter points, with the notation blue (0$<Br(Z\rightarrow e{\tau})<7\times10^{-15}$), yellow ($7\times10^{-15}\leq Br(Z\rightarrow e{\tau})<1\times10^{-14}$) , green ($1\times10^{-14}$ $\leq Br(Z\rightarrow e{\tau})<2\times10^{-14}$) and red ($2\times10^{-14}$ $\leq Br(Z\rightarrow e{\tau})<9.8\times10^{-6}$).}
{\label {N8}}
\end{figure}

 Next, supposing $M_S=1.2~{\rm TeV}$, we randomly scan the parameters. We scatter points on $Z\rightarrow e{\tau}$ in Fig.\ref {N8}. Some parameters ranges of $\tan\beta$, $M_1$, $M_2$, $M^2_{\tilde{L}ii}$, $M^2_{\tilde{E}ii}$, $M^2_{{\tilde{\nu}}ii}$, $T_{eii}$ and $T_{{\nu}ii}$ (i=1,2,3)  are given in the Table \ref {V}. In addition, other parameter spaces are also represented in the Table \ref {VI}. We use blue (0$<Br(Z\rightarrow e{\tau})<7\times10^{-15}$), yellow ($7\times10^{-15}\leq Br(Z\rightarrow e{\tau})<1\times10^{-14}$) , green ($1\times10^{-14}$ $\leq Br(Z\rightarrow e{\tau})<2\times10^{-14}$) and red ($2\times10^{-14}$ $\leq Br(Z\rightarrow e{\tau})<9.8\times10^{-6}$) to represent the results in different parameter spaces for the process of $Z\rightarrow e{\tau}$.

\begin{table*}
\caption{Scanning parameters for Fig.{\ref {N8}} and Fig.{\ref {N13}}}\label{VI}
\begin{tabular}{|c|c|c|}
\hline
Parameters&Min&Max\\
\hline
$\hspace{1.5cm}M^2_{\tilde{L}13}/\rm GeV^2\hspace{1.5cm}$ &$\hspace{1.5cm}0\hspace{1.5cm}$& $\hspace{1.5cm}10^6\hspace{1.5cm}$\\
\hline
$\hspace{1.5cm}M^2_{\tilde{E}13}/\rm GeV^2\hspace{1.5cm}$ &$\hspace{1.5cm}0\hspace{1.5cm}$& $\hspace{1.5cm}10^6\hspace{1.5cm}$\\
\hline
$\hspace{1.5cm}M^2_{\tilde{\nu}13}/\rm GeV^2\hspace{1.5cm}$ &$\hspace{1.5cm}0\hspace{1.5cm}$ &$\hspace{1.5cm}10^6\hspace{1.5cm}$\\
\hline
$T_{e13}$/GeV & $\hspace{1.5cm}-300\hspace{1.5cm}$ &$\hspace{1.5cm}300\hspace{1.5cm}$\\
\hline
$T_{{\nu}13}$/GeV &$\hspace{1.5cm}-500\hspace{1.5cm}$ &$\hspace{1.5cm}500\hspace{1.5cm}$\\
\hline
\end{tabular}
\end{table*}

 The analysis of the relationship between $M^2_{\tilde{L}13}$ and $M^2_{\tilde{E}13}$ is shown in Fig.\ref {N8}.
All the points are arranged in an arc in Fig.\ref {N8}. We can see that the overall change trend of scattered points is obvious, where four types of points are concentrated in $0~{\rm GeV}^2<M^2_{\tilde{E}13}<3\times10^{5}~{\rm GeV}^2$.
Blue parts are mainly in  $0~{\rm GeV}^2<M^2_{\tilde{L}13}<2.2\times10^{5}~{\rm GeV}^2$ and $0~{\rm GeV}^2<M^2_{\tilde{E}13}<1.25\times10^{5}~{\rm GeV}^2$, yellow parts are mainly  in  $2.2\times10^{5}~{\rm GeV}^2<M^2_{\tilde{L}13}<2.6\times10^{5}~{\rm GeV}^2$ and $1.25\times10^{5}~{\rm GeV}^2<M^2_{\tilde{E}13}<1.5\times10^{5}~{\rm GeV}^2$, green parts are mainly in  $2.6\times10^{5}~{\rm GeV}^2<M^2_{\tilde{L}13}<3.6\times10^{5}~{\rm GeV}^2$ and $1.5\times10^{5}~{\rm GeV}^2<M^2_{\tilde{E}13}<2.1\times10^{5}~{\rm GeV}^2$ and red parts are mainly  in  $3.6\times10^{5}~{\rm GeV}^2<M^2_{\tilde{L}13}<4.6\times10^{5}~{\rm GeV}^2$ and $2.1\times10^{5}~{\rm GeV}^2<M^2_{\tilde{E}13}<3\times10^{5}~{\rm GeV}^2$.

\subsection{$Z\rightarrow {\mu}{\tau}$}

The experimental upper bound for the LFV process $Z\rightarrow {\mu}{\tau}$ is $1.2\times10^{-5}$, which is about one order of
magnitude larger than the process $Z\rightarrow e{\mu}$.
The contributions from neutralino-slepton and chargino-sneutrino can be influenced by the parameters $M^2_{\tilde{E}23}$ , $M^2_{\tilde{L}23}$ and $T_{{\nu}23}$. Through experimental analysis, we can find that the law of $Z\rightarrow {\mu}{\tau}$ process is similar to those of $Z\rightarrow e{\mu}$ process and $Z\rightarrow e{\tau}$ process.
The branching ratios increase with the increase of variables $M^2_{\tilde{E}23}$, $M^2_{\tilde{L}23}$ and $T_{{\nu}23}$. When $M^2_{\tilde{E}23}$ is the variable, the branching ratio of $Z\rightarrow {\mu}{\tau}$ can reach $10^{-13}$. When $M^2_{\tilde{L}23}$ is the variable, the branching ratio of $Z\rightarrow {\mu}{\tau}$ can reach $10^{-11}$. When $T_{{\nu}23}$ is the variable, the branching ratio of $Z\rightarrow {\mu}{\tau}$ can reach $10^{-9}$. It can be deduced that parameter $T_{{\nu}23}$ is more sensitive than parameter $M^2_{\tilde{E}23}$ and more sensitive than parameter $M^2_{\tilde{L}23}$.

\begin{figure}[h]
\setlength{\unitlength}{5mm}
\centering
\includegraphics[width=3.0in]{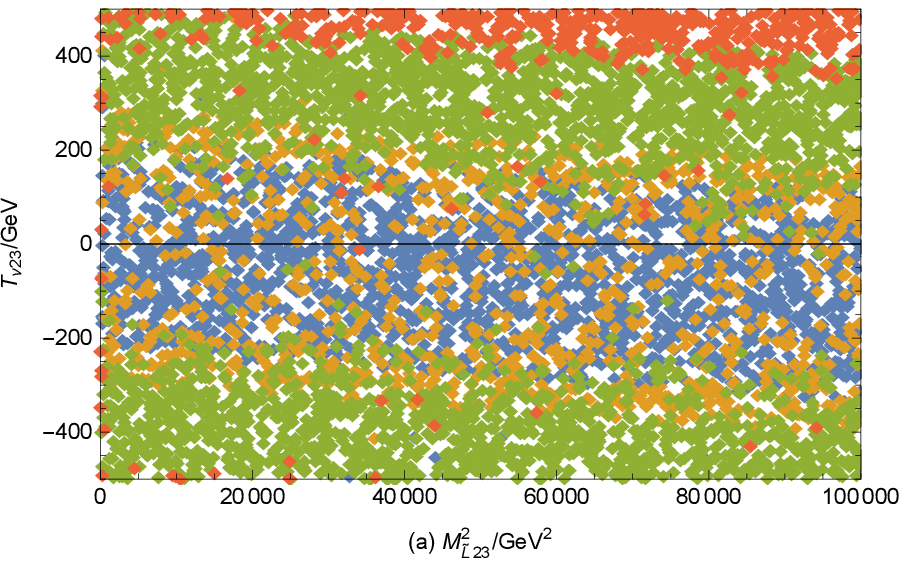}
\vspace{0.2cm}
\setlength{\unitlength}{5mm}
\centering
\includegraphics[width=3.0in]{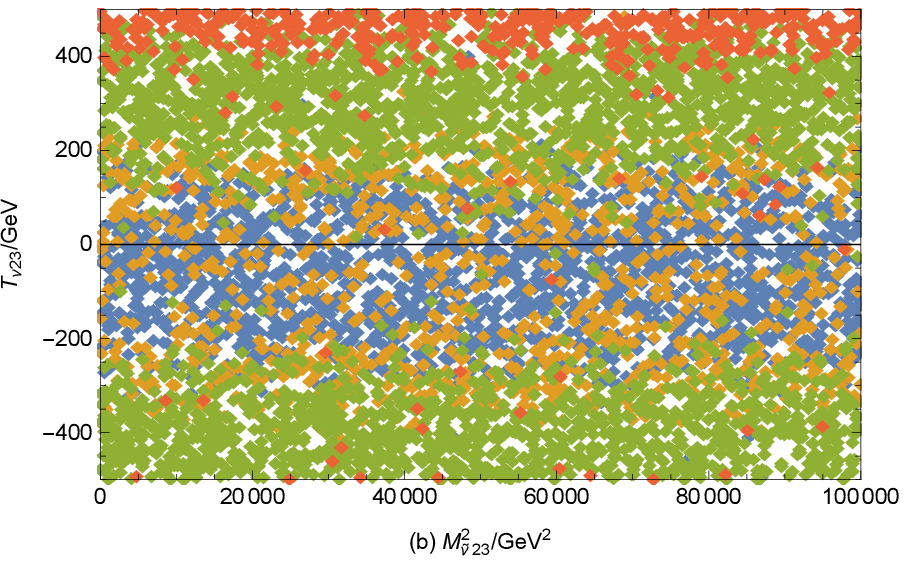}
\vspace{0.2cm}
\setlength{\unitlength}{5mm}
\centering
\includegraphics[width=3.0in]{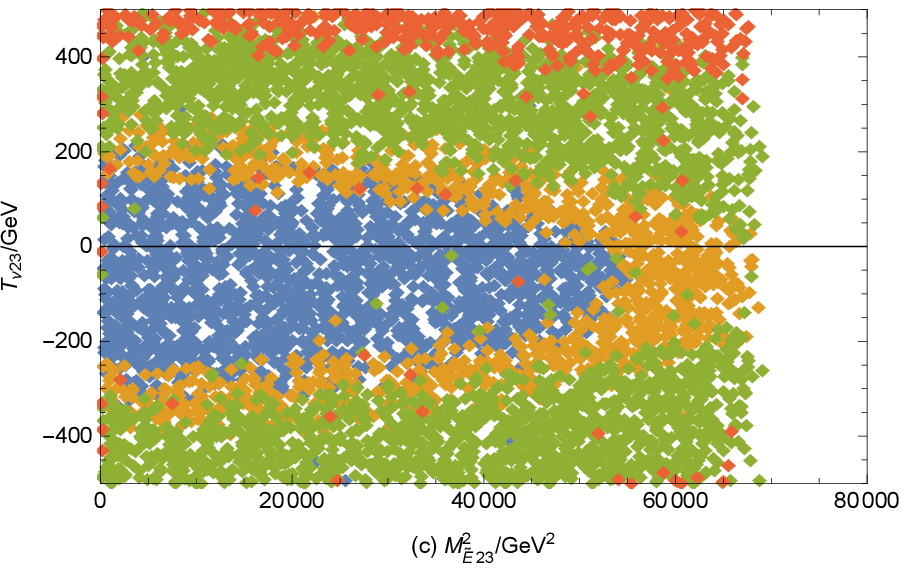}
\caption{Under the premise of current limit on lepton flavor violating decay $Z\rightarrow {\mu}{\tau}$. Reasonable parameter space is selected to scatter points, with the notation blue (0$<Br(Z\rightarrow {\mu}{\tau})<6\times10^{-13}$), yellow ($6\times10^{-13}\leq Br(Z\rightarrow {\mu}{\tau})<10^{-12}$), green ($10^{-12}$ $\leq Br(Z\rightarrow {\mu}{\tau})<3\times10^{-12}$) and red ($3\times10^{-12}$ $\leq Br(Z\rightarrow {\mu}{\tau})<1.2\times10^{-5}$).}\label{N9}
\end{figure}

 Next, we scatter points on $Z\rightarrow {\mu}{\tau}$ in Fig.\ref {N9} with the parameters in the Table \ref {VII}. These points are divided into blue (0$<Br(Z\rightarrow {\mu}{\tau})<6\times10^{-13}$), yellow ($6\times10^{-13}\leq Br(Z\rightarrow {\mu}{\tau})<10^{-12}$), green ($10^{-12}$ $\leq Br(Z\rightarrow {\mu}{\tau})<3\times10^{-12}$) and red ($3\times10^{-12}$ $\leq Br(Z\rightarrow {\mu}{\tau})<1.2\times10^{-5}$) to represent the results in different parameter spaces for the process of $Z\rightarrow {\mu}{\tau}$.

\begin{table*}
\caption{Scanning parameters for Fig.{\ref {N9}} with i=1,2,3}\label{VII}
\begin{tabular}{|c|c|c|}
\hline
Parameters&Min&Max\\
\hline
$\hspace{1.5cm}M^2_{\tilde{L}23}/\rm GeV^2\hspace{1.5cm}$ &$\hspace{1.5cm}0\hspace{1.5cm}$& $\hspace{1.5cm}10^5\hspace{1.5cm}$\\
\hline
$\hspace{1.5cm}M^2_{\tilde{E}23}/\rm GeV^2\hspace{1.5cm}$ &$\hspace{1.5cm}0\hspace{1.5cm}$& $\hspace{1.5cm}10^5\hspace{1.5cm}$\\
\hline
$\hspace{1.5cm}M^2_{\tilde{\nu}23}/\rm GeV^2\hspace{1.5cm}$ &$\hspace{1.5cm}0\hspace{1.5cm}$ &$\hspace{1.5cm}10^5\hspace{1.5cm}$\\
\hline
$T_{e23}$/GeV & $\hspace{1.5cm}-300\hspace{1.5cm}$ &$\hspace{1.5cm}300\hspace{1.5cm}$\\
\hline
$T_{{\nu}23}$/GeV &$\hspace{1.5cm}-500\hspace{1.5cm}$ &$\hspace{1.5cm}500\hspace{1.5cm}$\\
\hline
$\hspace{1.5cm}M^2_{\tilde{L}ii}/\rm GeV^2\hspace{1.5cm}$ &$\hspace{1.5cm}2\times10^{5}\hspace{1.5cm}$& $\hspace{1.5cm}10^8\hspace{1.5cm}$\\
\hline
$\hspace{1.5cm}M^2_{\tilde{E}ii}/\rm GeV^2\hspace{1.5cm}$ &$\hspace{1.5cm}2\times10^{5}\hspace{1.5cm}$& $\hspace{1.5cm}10^8\hspace{1.5cm}$\\
\hline
$\hspace{1.5cm}M^2_{\tilde{\nu}ii}/\rm GeV^2\hspace{1.5cm}$ &$\hspace{1.5cm}1\times10^{5}\hspace{1.5cm}$ &$\hspace{1.5cm}10^8\hspace{1.5cm}$\\
\hline
$T_{eii}$/GeV & $\hspace{1.5cm}-3000\hspace{1.5cm}$ &$\hspace{1.5cm}3000\hspace{1.5cm}$\\
\hline
$T_{{\nu}ii}$/GeV &$\hspace{1.5cm}-3000\hspace{1.5cm}$ &$\hspace{1.5cm}3000\hspace{1.5cm}$\\
\hline
\end{tabular}
\end{table*}

 The analysis of the relationship between $M^2_{\tilde{L}23}$ and $T_{{\nu}23}$,~$M^2_{\tilde{\nu}23}$ and $T_{{\nu}23}$,~$M^2_{\tilde{E}23}$ and $T_{{\nu}23}$ are shown in Fig.\ref {N9}.
In Fig.\ref {N9}, we can see four of these points concentrated in -500 GeV$<T_{{\nu}23}<500$ GeV. In Fig.\ref {N9}(a), blue parts are mainly in -300 GeV$<T_{{\nu}23}<180$ GeV, yellow parts are mainly  in -400 GeV$<T_{{\nu}23}<200$ GeV, green parts are mainly in -500 GeV$<T_{{\nu}23}<-200$ GeV and 200 GeV$<T_{{\nu}23}<400$ GeV and red parts are mainly  in 400 GeV$<T_{{\nu}23}<500$ GeV.
In Fig.\ref {N9}(b), blue parts are mainly in -200 GeV$<T_{{\nu}23}<180$ GeV, yellow parts are mainly  in -400 GeV$<T_{{\nu}23}<200$ GeV, green parts are mainly in -500 GeV$<T_{{\nu}23}<-200$ GeV and 200 GeV$<T_{{\nu}23}<400$ GeV and red parts are mainly  in 400 GeV$<T_{{\nu}23}<500$ GeV. We find that the change trends in Fig.\ref {N9}(a) and Fig.\ref {N9}(b) are relatively weak, but Fig.\ref {N9}(a) is relatively obvious than Fig.\ref {N9}(b).

 Finally, we analyze the effect from parameters $M^2_{\tilde{E}23}$ and $T_{{\nu}23}$ in Fig.\ref {N9}(c).
Blue parts are almost symmetrically distributed about $T_{{\nu}23}=0$ and concentrate in the -200 GeV$<T_{{\nu}23}<200$ GeV and $0~{\rm GeV}^2<M^2_{\tilde{E}23}<5.5\times10^{4}~{\rm GeV}^2$. Yellow parts are mainly distributed outside green parts.
Green parts are mainly in -500 GeV$<T_{{\nu}23}<-300$ GeV and 200 GeV$<T_{{\nu}23}<400$ GeV and red parts are mainly  in 400 GeV$<T_{{\nu}23}<500$ GeV.

\subsection{$h\rightarrow e{\mu}$}
\begin{figure}[h]
\setlength{\unitlength}{5mm}
\centering
\includegraphics[width=3.0in]{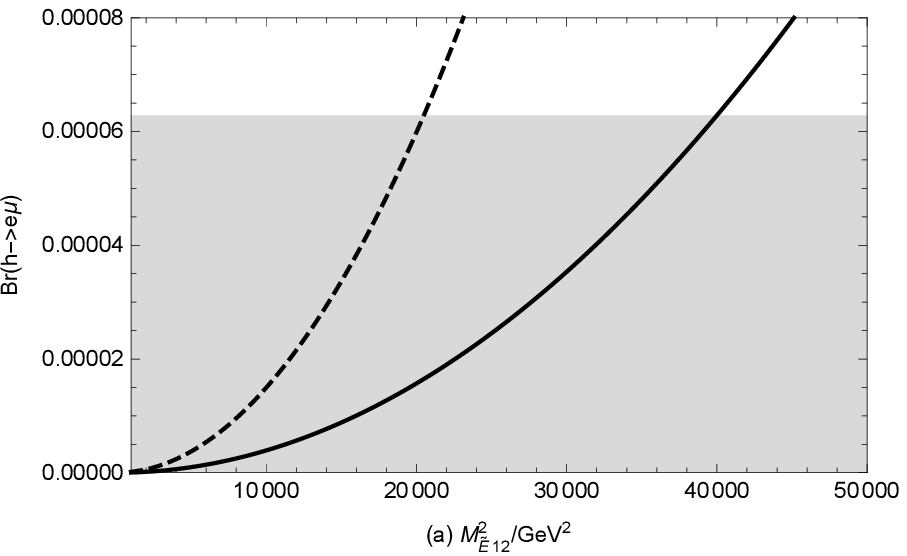}
\vspace{0.2cm}
\setlength{\unitlength}{5mm}
\centering
\includegraphics[width=3.0in]{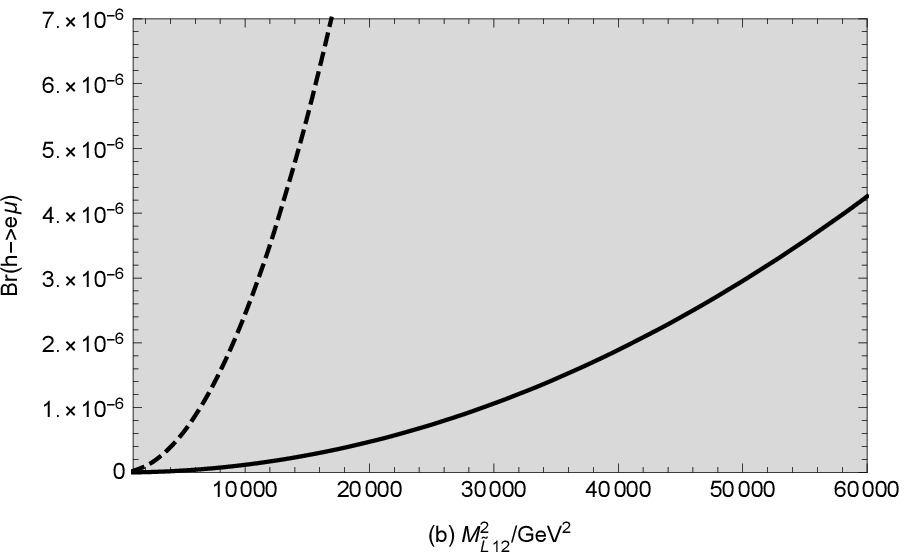}
\vspace{0.2cm}
\setlength{\unitlength}{5mm}
\centering
\includegraphics[width=3.0in]{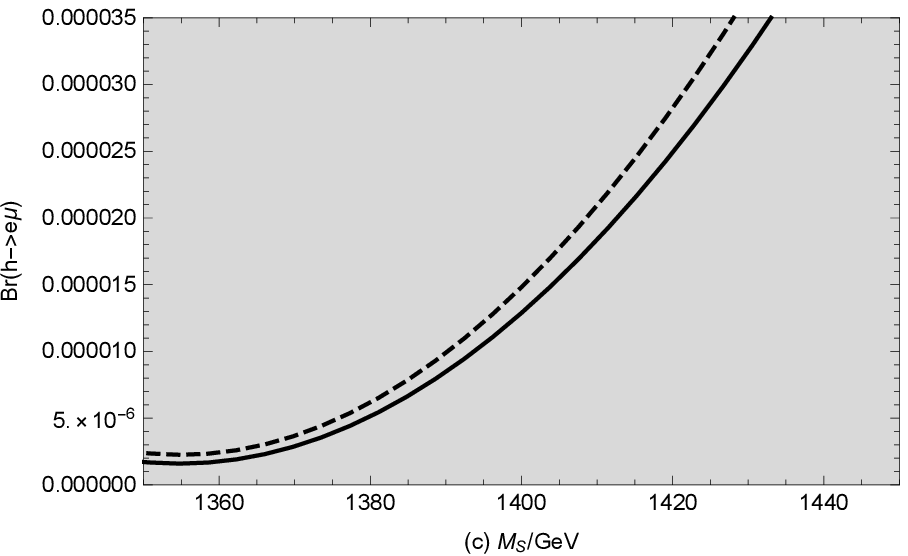}
\caption{$Br(h\rightarrow e{\mu})$ schematic diagrams affected by different parameters. The gray area is reasonable value range, where $Br(h\rightarrow e{\mu})$ satisfies the upper limit. The dashed and solid lines in Fig.\ref {N10}(a)(b) correspond to $M_S =1.5~{\rm TeV}$ and $M_S =1.2~{\rm TeV}$. The dashed and solid lines in Fig.\ref {N10}(c) correspond to $M^2_{\tilde{L}12}=6\times10^{3}~{\rm GeV}^2$ and $M^2_{\tilde{L}12}=5\times10^{3}~{\rm GeV}^2$.}\label{N10}
\end{figure}

In this subsection, we mainly analyze 125 GeV Higgs boson decays with LFV $h\rightarrow e{\mu}$ in the $U(1)_X$SSM.
With the same parameters as $Z\rightarrow e{\mu}$ process, we paint $Br(h\rightarrow e{\mu})$ schematic diagrams affected by different parameters in the Fig.\ref {N10}.

 In the Fig.\ref {N10}(a), we plot $Br(h\rightarrow e{\mu})$ versus $M^2_{\tilde{E}12}$, in which the numerical results are shown by the dashed curve and solid curve corresponding
to $M_S =1.5~{\rm TeV}$ and $M_S =1.2~{\rm TeV}$ respectively. $Br(h\rightarrow e{\mu})$ varies with $M^2_{\tilde{E}12}$ in the range from $0~{\rm GeV}^2$ to $5\times10^{4}~{\rm GeV}^2$. It can be clearly seen that both the solid line and the dashed line have an upward trend. The rising trend of the dashed line is greater than that of the solid line.
Gray region represents the experimental limit.
The dashed line in the range of $0~{\rm GeV}^2$ to $2\times10^{4}~{\rm GeV}^2$ and solid line in the range of
$0~{\rm GeV}^2$ to $4\times10^{4}~{\rm GeV}^2$ are in the gray area.

 In the Fig.\ref {N10}(b), we plot $Br(h\rightarrow e{\mu})$ versus $M^2_{\tilde{L}12}$, in which the dashed curve corresponds to $M_S =1.5~{\rm TeV}$ and the solid line corresponds to
$M_S =1.2~{\rm TeV}$. We can clearly see that the dashed line increases with the increasing $M^2_{\tilde{L}12}$ in the range of $0~{\rm GeV}^2$ to $2\times10^{4}~{\rm GeV}^2$. The solid line increases with the increasing $M^2_{\tilde{L}12}$ in the range of $0~{\rm GeV}^2$ to $5\times10^{4}~{\rm GeV}^2$. The dashed curve is also larger than the solid curve. The solid line and the dashed line are located in the gray area. So, the contributions can be influenced obviously by the parameters $M^2_{\tilde{E}12}$ and $M^2_{\tilde{L}12}$.

$Br(h\rightarrow e{\mu})$ versus $M_S$ is plotted in the Fig.\ref {N10}(c), where
 the dashed curve corresponds to $M^2_{\tilde{L}12}=6\times10^{3}~{\rm GeV}^2$ and the solid line corresponds to
$M^2_{\tilde{L}12}=5\times10^{3}~{\rm GeV}^2$. It is clear that both the dashed and the solid
line are the increasing functions of $M_S$ in the range of $1350~{\rm GeV}$ to $1450~{\rm GeV}$.
The dashed curve is also larger than the solid curve. The solid line and the dashed line reach $3.0\times10^{-5}$ and are located in the gray area.
If $M^2_{\tilde{E}12}$ and $M^2_{\tilde{L}12}$ are very small, the $Br(h\rightarrow e{\mu})$ turns to small quickly, and the reasonable range of $M_S$ becomes large.

\begin{figure}[h]
\setlength{\unitlength}{5mm}
\centering
\includegraphics[width=3.0in]{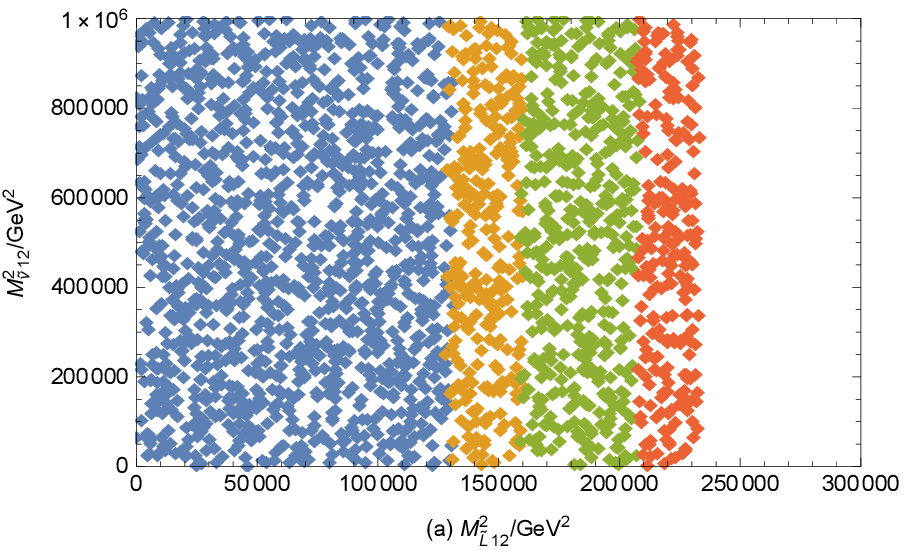}
\vspace{0.2cm}
\setlength{\unitlength}{5mm}
\centering
\includegraphics[width=3.0in]{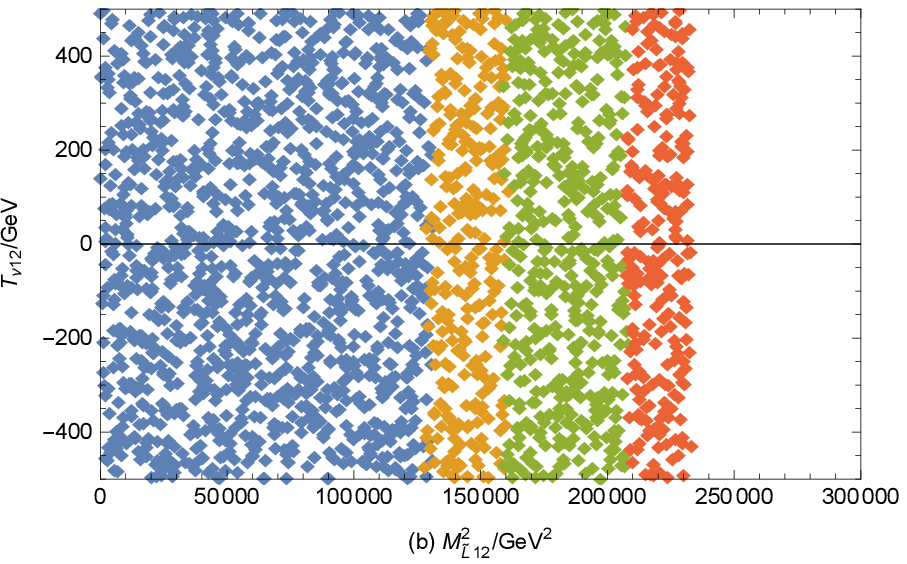}
\caption{Under the premise of current limit on lepton flavor violating decay $h\rightarrow e{\mu}$. Reasonable parameter space is selected to scatter points, with the notation blue (0$<Br(h\rightarrow e{\mu})<2\times10^{-5}$), yellow ($2\times10^{-5}\leq Br(h\rightarrow e{\mu})<3\times10^{-5}$) , green ($3\times10^{-5}$ $\leq Br(h\rightarrow e{\mu})<5\times10^{-5}$) and red ($5\times10^{-5}$ $\leq Br(h\rightarrow e{\mu})<6.28\times10^{-5}$).}
{\label {N11}}
\end{figure}

 Supposing $M_S=1.2~{\rm TeV}$, $M_1=1.2~{\rm TeV}$, we randomly scan the parameters. These parameter ranges are given in the Table \ref {V} and Fig.\ref {N11} is obtained. We use blue (0$<Br(h\rightarrow e{\mu})<2\times10^{-5}$), yellow ($2\times10^{-5}\leq Br(h\rightarrow e{\mu})<3\times10^{-5}$) , green ($3\times10^{-5}$ $\leq Br(h\rightarrow e{\mu})<5\times10^{-5}$) and red ($5\times10^{-5}$ $\leq Br(h\rightarrow e{\mu})<6.28\times10^{-5}$) to represent the results in different parameter spaces for the process of $Z\rightarrow e{\mu}$.

 The relationship between $M^2_{\tilde{L}12}$ and $M^2_{{\nu}12}$ is shown in Fig.\ref {N11}(a). The relationship between $M^2_{\tilde{L}12}$ and $T_{{\nu}12}$ is shown
in Fig.\ref {N11}(b). All points are clearly distributed in their respective regions. The four types of points are concentrated in
$0~{\rm GeV}^2<M^2_{\tilde{L}12}<2.3\times10^{5}~{\rm GeV}^2$. Blue parts are mainly in $0~{\rm GeV}^2<M^2_{\tilde{E}12}<1.3\times10^{5}~{\rm GeV}^2$, yellow parts are mainly  in $1.3\times10^{5}~{\rm GeV}^2<M^2_{\tilde{E}12}<1.6\times10^{5}~{\rm GeV}^2$, green parts are mainly
in $1.6\times10^{5}~{\rm GeV}^2<M^2_{\tilde{E}12}<2.1\times10^{5}~{\rm GeV}^2$ GeV and red parts are mainly in $2.1\times10^{5}~{\rm GeV}^2<M^2_{\tilde{E}12}<2.3\times10^{5}~{\rm GeV}^2$.

\subsection{$h\rightarrow e{\tau}$}
\begin{figure}[h]
\setlength{\unitlength}{5mm}
\centering
\includegraphics[width=3.0in]{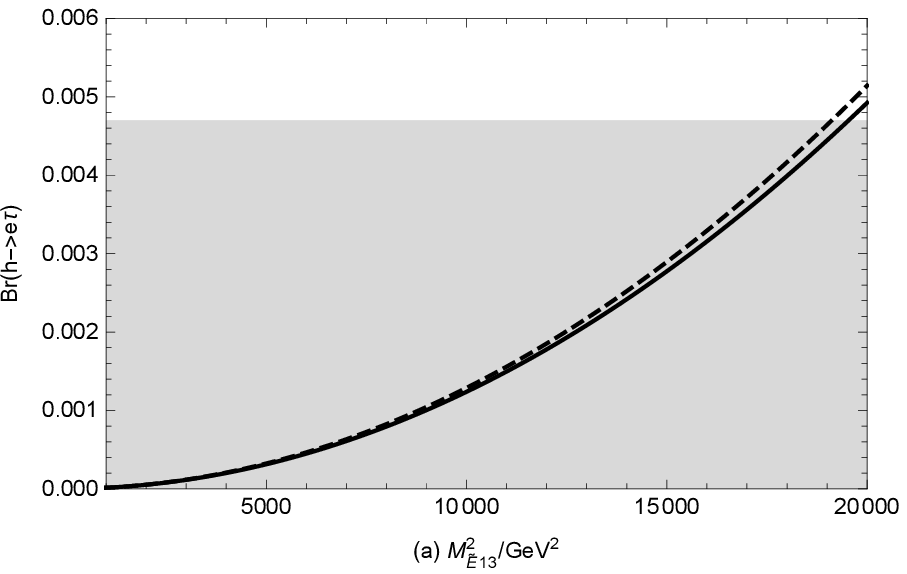}
\vspace{0.2cm}
\setlength{\unitlength}{5mm}
\centering
\includegraphics[width=3.0in]{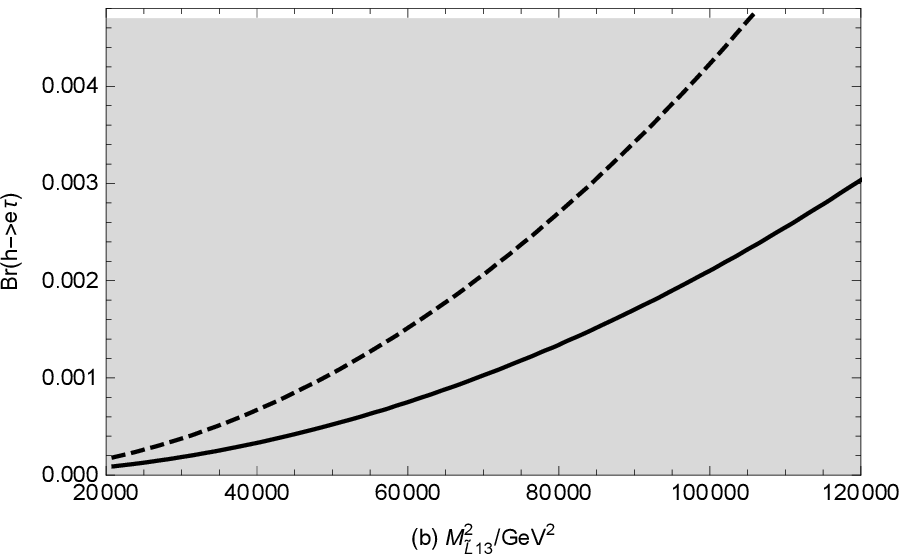}
\caption{$Br(h\rightarrow e{\tau})$ schematic diagrams affected by different parameters. The gray area is reasonable value range, where $Br(Z\rightarrow e{\tau})$ satisfies the upper limit. The dashed and solid lines in Fig.\ref {N12}(a)(b) correspond to $M^2_{\tilde{L}ii}=2.5\times10^{6}~{\rm GeV}^2$ and $M^2_{\tilde{L}ii}=3\times10^{6}~{\rm GeV}^2$ with  i=1,2,3.}\label{N12}
\end{figure}

In this section, we analyze the 125 GeV Higgs boson decay $h\rightarrow e{\tau}$ in the $U(1)_X$SSM model.
With the same parameters as $Z\rightarrow e{\tau}$ process, we paint $Br(h\rightarrow e{\tau})$ schematic diagrams affected by different parameters in the Fig.\ref {N12}.

 Setting $v_S=4.3~{\rm TeV}$, we plot $Br(h\rightarrow e{\tau})$ versus $M^2_{\tilde{E}13}$ in the Fig.\ref {N12}(a).
The dashed curve corresponds to $M^2_{\tilde{L}ii}=2.5\times10^{6}~{\rm GeV}^2$ and the solid line corresponds to $M^2_{\tilde{L}ii}=3\times10^{6}~{\rm GeV}^2$, (i=1,2,3).
We can clearly see that the two lines increase with the increasing $M^2_{\tilde{E}13}$ in the range of $0~{\rm GeV}^2-2\times10^{4}~{\rm GeV}^2$. The dashed curve is larger than the solid curve. The solid line part of $0~{\rm GeV}^2<M^2_{\tilde{E}13}<1.96\times10^{4}~{\rm GeV}^2$ is in the gray area and the dashed line part of $0~{\rm GeV}^2<M^2_{\tilde{E}13}<1.9\times10^{4}~{\rm GeV}^2$ is in the gray area. That is to say the dashed line and the solid line of the rest exceed the gray area.

 In the Fig.\ref {N12}(b), we show $h\rightarrow e{\tau}$ versus $M^2_{\tilde{L}13}$, where the dashed curve corresponds to $M^2_{\tilde{L}ii}=2.5\times10^{6}~{\rm GeV}^2$ and the solid line corresponds to $M^2_{\tilde{L}ii}=3\times10^{6}~{\rm GeV}^2$, (i=1,2,3). During the $2\times10^{3}~{\rm GeV}^2<M^2_{\tilde{L}13}<1.2\times10^{5}~{\rm GeV}^2$, both the dashed line and the solid line are increasing functions, and the slope of the dashed line is greater than that of the solid line. The solid line part as a whole and the dashed line part of $2\times10^{3}~{\rm GeV}^2<M^2_{\tilde{L}13}<1.05\times10^{5}~{\rm GeV}^2$ are in the gray area and the dashed line of the rest exceeds the gray area. The contributions can be influenced obviously by the parameters $M^2_{\tilde{E}13}$ and $M^2_{\tilde{L}13}$.

\begin{figure}[h]
\setlength{\unitlength}{5mm}
\centering
\includegraphics[width=3.0in]{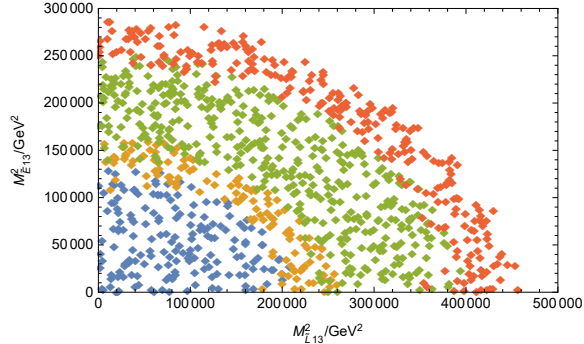}
\caption{Under the premise of current limit on lepton flavor violating decay $h\rightarrow e{\tau}$. Reasonable parameter space is selected to scatter points, with the notation blue (0$<Br(h\rightarrow e{\tau})<10^{-3}$), yellow ($10^{-3}\leq Br(h\rightarrow e{\tau})<1.5\times10^{-3}$), green ($1.5\times10^{-3}$ $\leq Br(h\rightarrow e{\tau})<3.5\times10^{-3}$) and red ($3.5\times10^{-3}$ $\leq Br(h\rightarrow e{\tau})<4.7\times10^{-3}$).}
{\label {N13}}
\end{figure}

 Next, supposing the parameters with $M_S=1.2~{\rm TeV}$, we randomly scan the parameters. We scatter points for $h\rightarrow e{\tau}$ in Fig.\ref {N13}. Some parameters ranges of $\tan\beta$, $M_1$, $M_2$, $M^2_{\tilde{L}ii}$, $M^2_{\tilde{E}ii}$, $M^2_{{\tilde{\nu}}ii}$, $T_{eii}$ and $T_{{\nu}ii}$ (i=1,2,3) are given in the Table \ref {V}. In addition, other parameter spaces are also represented in the Table \ref {VI}. We use blue (0$<Br(h\rightarrow e{\tau})<10^{-3}$), yellow ($10^{-3}\leq Br(h\rightarrow e{\tau})<1.5\times10^{-3}$), green ($1.5\times10^{-3}$ $\leq Br(h\rightarrow e{\tau})<3.5\times10^{-3}$) and red ($3.5\times10^{-3}$ $\leq Br(h\rightarrow e{\tau})<4.7\times10^{-3}$) to represent the results in different parameter spaces for the process of $h\rightarrow e{\tau}$.

 Finally, we analysize of the relationship between $M^2_{\tilde{L}13}$ and $M^2_{\tilde{E}13}$ in Fig.\ref {N13}.
All scatters are fan-shaped evenly distributed, where four types of points are concentrated in $0~{\rm GeV}^2<M^2_{\tilde{L}13}<4.6\times10^{5}~{\rm GeV}^2$.
Blue parts are mainly in  $0~{\rm GeV}^2<M^2_{\tilde{L}13}<2\times10^{5}~{\rm GeV}^2$ and $0~{\rm GeV}^2<M^2_{\tilde{E}13}<1.3\times10^{5}~{\rm GeV}^2$, yellow parts are mainly  in  $2\times10^{5}~{\rm GeV}^2<M^2_{\tilde{L}13}<2.6\times10^{5}~{\rm GeV}^2$ and $1.3\times10^{5}~{\rm GeV}^2<M^2_{\tilde{E}13}<1.6\times10^{5}~{\rm GeV}^2$, green parts are mainly in  $2.6\times10^{5}~{\rm GeV}^2<M^2_{\tilde{L}13}<4\times10^{5}~{\rm GeV}^2$ and $1.6\times10^{5}~{\rm GeV}^2<M^2_{\tilde{E}13}<2.5\times10^{5}~{\rm GeV}^2$ and red parts are mainly  in  $4\times10^{5}~{\rm GeV}^2<M^2_{\tilde{L}13}<4.6\times10^{5}~{\rm GeV}^2$ and $2.5\times10^{5}~{\rm GeV}^2<M^2_{\tilde{E}13}<2.9\times10^{5}~{\rm GeV}^2$.

\subsection{$h\rightarrow {\mu}{\tau}$}
 At the last, we analyze the process $h\rightarrow {\mu}{\tau}$ in the $U(1)_X$SSM. After experimental exploration, the experimental law of $h\rightarrow {\mu}{\tau}$ process is similar to those of $h\rightarrow e{\mu}$ process and $h\rightarrow e{\tau}$ process. When $M^2_{\tilde{E}23}$, $M^2_{\tilde{L}23}$, $T_{{\nu}23}$ are variables, the corresponding branching ratios can reach $10^{-4}$, $10^{-4}$, $10^{-9}$ respectively.  It can be deduced that parameters $M^2_{\tilde{E}23}$ and $M^2_{\tilde{L}23}$ more sensitive than the parameter $T_{{\nu}23}$.

\begin{figure}[h]
\setlength{\unitlength}{5mm}
\centering
\includegraphics[width=3.0in]{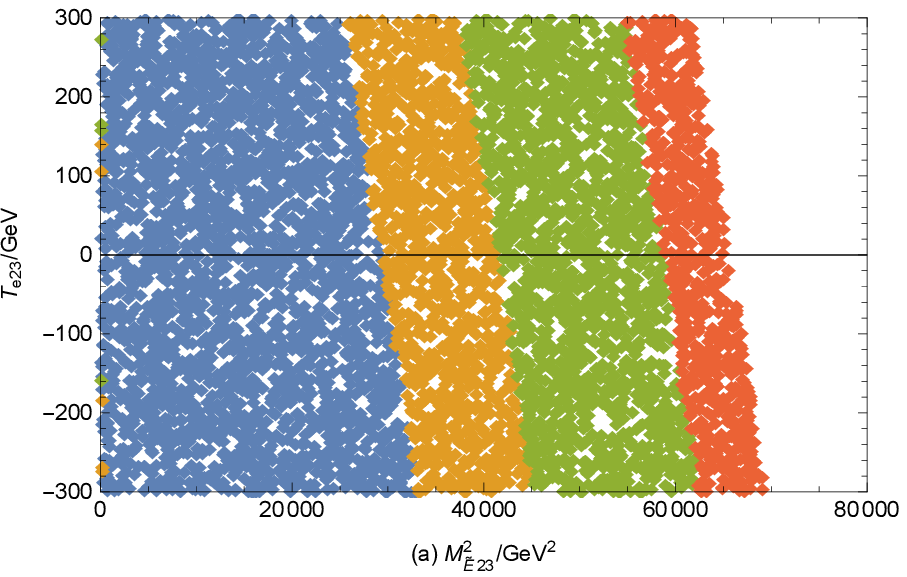}
\vspace{0.2cm}
\setlength{\unitlength}{5mm}
\centering
\includegraphics[width=3.0in]{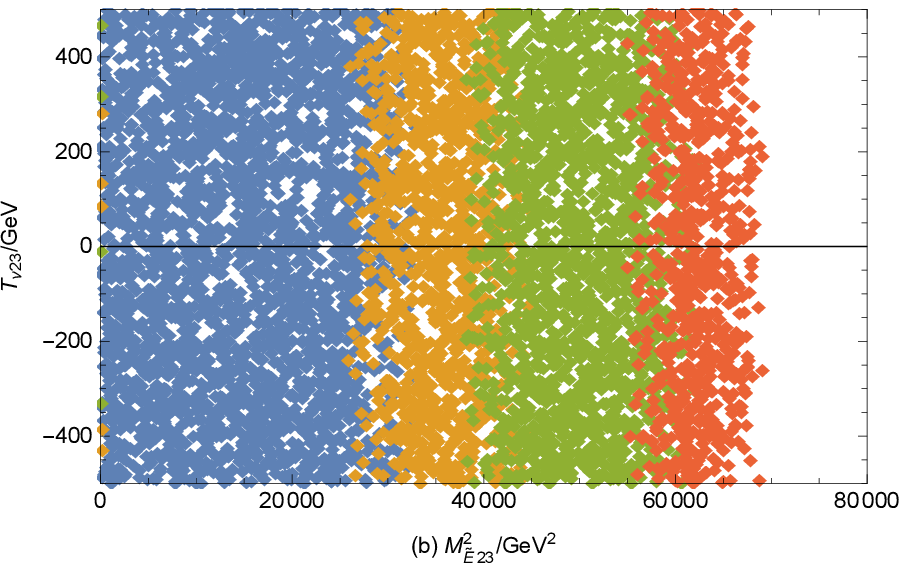}
\caption{Under the premise of current limit on lepton flavor violating decay $h\rightarrow {\mu}{\tau}$. Reasonable parameter space is selected to scatter points, with the notation blue (0$<Br(h\rightarrow {\mu}{\tau})<5\times10^{-4}$), yellow ($5\times10^{-4}\leq Br(h\rightarrow {\mu}{\tau})<1\times10^{-3}$) , green ($1\times10^{-3}$ $\leq Br(h\rightarrow {\mu}{\tau})<2\times10^{-3}$) and red ($2\times10^{-3}$ $\leq Br(h\rightarrow {\mu}{\tau})<2.5\times10^{-3}$).}\label{N14}
\end{figure}

 Next, we scatter points on $h\rightarrow {\mu}{\tau}$ in Fig.\ref {N14} with the parameters in the Table \ref {VII}. These points are divided into blue (0$<Br(h\rightarrow {\mu}{\tau})<5\times10^{-4}$), yellow ($5\times10^{-4}\leq Br(h\rightarrow {\mu}{\tau})<1\times10^{-3}$), green ($1\times10^{-3}$ $\leq Br(h\rightarrow {\mu}{\tau})<2\times10^{-3}$) and red ($2\times10^{-3}$ $\leq Br(h\rightarrow {\mu}{\tau})<2.5\times10^{-3}$) to represent the results in different parameter spaces for the process of $h\rightarrow e{\tau}$.

 We plot $M^2_{\tilde{E}23}$ varying with $T_{{e}23}$ in Fig.\ref{N14}(a), where we can see four of these points concentrated in $0~{\rm GeV}^2<M^2_{\tilde{E}23}<7\times10^{4}~{\rm GeV}^2$. In Fig.\ref {N14}(a), blue parts are mainly in $0~{\rm GeV}^2<M^2_{\tilde{E}23}<3\times10^{4}~{\rm GeV}^2$, yellow parts are mainly
in $3\times10^{4}~{\rm GeV}^2<M^2_{\tilde{E}23}<4.5\times10^{4}~{\rm GeV}^2$, green parts are mainly in $4.5\times10^{4}~{\rm GeV}^2<M^2_{\tilde{E}23}<6\times10^{4}~{\rm GeV}^2$  and red parts are mainly  in $6\times10^{4}~{\rm GeV}^2<M^2_{\tilde{E}23}<7\times10^{4}~{\rm GeV}^2$. In Fig.\ref {N14}(b), we can find that $T_{{\nu}23}$ is not sensitive. Blue parts are mainly in $0~{\rm GeV}^2<M^2_{\tilde{E}23}<2.7\times10^{4}~{\rm GeV}^2$, yellow parts are mainly in $2.7\times10^{4}~{\rm GeV}^2<M^2_{\tilde{E}23}<4\times10^{4}~{\rm GeV}^2$, green parts are mainly in $4\times10^{4}~{\rm GeV}^2<M^2_{\tilde{E}23}<6\times10^{4}~{\rm GeV}^2$  and red parts are mainly in $6\times10^{4}~{\rm GeV}^2<M^2_{\tilde{E}23}<7\times10^{4}~{\rm GeV}^2$. We can clearly see that the change trend of Fig.\ref {N14}(a) is more obvious than that of Fig.\ref {N14}(b).

\section{discussion and conclusion}

 In this paper, we have studied the LFV processes $Z\rightarrow{{l_i}^{\pm}{l_j}^{\mp}}$ and $h\rightarrow{{l_i}^{\pm}{l_j}^{\mp}}$ in the $U(1)_X$SSM.
We take into account the one loop diagrams which include self-energy diagram, triangle diagram. In the numerical calculation, we
scan large parameter spaces and make a rich numerical results. In our used parameter space, the numerical results show that the rates for $Br(Z\rightarrow{{l_i}^{\pm}{l_j}^{\mp}})$ and $Br(h\rightarrow{{l_i}^{\pm}{l_j}^{\mp}})$
can almost reach their present experimental upper bounds. The numerical analyses indicate that $M_1$, $M_2$, $g_{YX}$, $\tan{\beta}$ are important parameters. The sensitive parameters are $M^2_{\tilde{E}ij}$, $M^2_{\tilde{L}ij}$, $M^2_{\tilde{\nu}ij}$, $T_{{\nu}ij}$ and $T_{{e}ij}$$(i\ne j)$, because they affect the results strongly. In the whole, the non-diagonal elements which correspond to the generations of the initial lepton and final lepton are main sensitive parameters and LFV sources. Most parameters can break the upper limit of the experiment and provide new ideas for finding NP.

 From the numerical results, the branching ratios of $Z\rightarrow e{\mu}$, $Z\rightarrow e{\tau}$, $Z\rightarrow {\mu}{\tau}$ and $h\rightarrow e{\mu}$,
$h\rightarrow e{\tau}$, $h\rightarrow {\mu}{\tau}$ depend on the slepton flavor mixing parameters. Through data analysis, we can get that the branching ratio of $Z\rightarrow e{\mu}$ can reach $10^{-11}$. The branching ratios of $Z\rightarrow e{\tau}$ and $Z\rightarrow {\mu}{\tau}$ can reach $10^{-9}$. The branching ratio of $h\rightarrow e{\mu}$ can reach $10^{-5}$. The branching ratios of $h\rightarrow e{\tau}$ and $h\rightarrow {\mu}{\tau}$ can reach $10^{-3}$. It is not difficult to find that the numerical results of process $h\rightarrow e{\tau}$ and process $h\rightarrow {\mu}{\tau}$ are very close, and the numerical results of process $Z\rightarrow e{\tau}$ and process $Z\rightarrow {\mu}{\tau}$ are very close. The branching ratios of $h\rightarrow e{\mu}$, $h\rightarrow e{\tau}$, $h\rightarrow {\mu}{\tau}$ in the $U(1)_X$SSM are close to the corresponding experimental upper limits
of $Br(h\rightarrow e{\mu})$, $Br(h\rightarrow e{\tau})$ and $Br(h\rightarrow {\mu}{\tau})$, which may be detected in the future.

The numerical study is performed in terms of the most relevant model parameters. It shows that the flavor mixing parameters (such as $M^2_{\tilde{E}12}$ and $M^2_{\tilde{L}12}$) are very important and will be most efficiently tested at LHC and the future colliders(such as CEPC/FCC-ee). At the future colliders, the more high statistic of Higgs boson and Z boson events will be achieved, and the sensitivities of the future colliders can be improved obviously. $M^2_{\tilde{E}12}$ and $M^2_{\tilde{L}12}$ are the core parameters for
$Z\rightarrow{e\mu}$ and $h\rightarrow{e\mu}$. Larger $M^2_{\tilde{E}12}$ and $M^2_{\tilde{L}12}$ can improve the branching ratios ($Br(Z\rightarrow{e\mu})~{\rm and}~Br(h\rightarrow{e\mu})$) obviously, and they are in the reachable region of LHC. As the typical parameter of the $U(1)_X$SSM, $M_S$ is the mass for
the super partner of the Higgs singlet $S$, which appears in the mass squared matrix of Higgs and the mass matrix of neutralino.
So, the search of $M_S$ should have relation with Higgs decays and the processes relating with neutralino. We hope that the researched LFV
decays $Z\rightarrow{{l_i}^{\pm}{l_j}^{\mp}}$ and $h\rightarrow{{l_i}^{\pm}{l_j}^{\mp}}$ can be detected in the LHC and future colliders.

{\bf Acknowledgments}

This work is supported by National Natural Science Foundation of China (NNSFC)
	(No. 11535002, No. 11705045), Natural Science Foundation of Hebei Province
	(A2020201002).

\appendix
\section{the couplings}

 The concrete forms of coupling coefficients corresponding to Fig.\ref{N1} are shown as:

 Fig.\ref{N1}(1):~$S_1=\widetilde{\nu}_n$, $S_2=\widetilde{\nu}_m$, $F=\chi^c$
\begin{eqnarray}\label{A1}
&&\hspace{3cm}H^{S_2F\overline{l}_i}_L(1)= -\frac{1}{\sqrt{2}}U_{\alpha2}^{*}Z_{mi}^{I,*}Y_{e,i},  <\frac{i}{\sqrt{2}}U_{\alpha2}^{*}Z_{mi}^{R,*}Y_{e,i}>,\nonumber\\
&&\hspace{3cm}H^{S_2F\overline{l}_i}_R(1)= \frac{1}{\sqrt{2}}g_2Z_{mi}^{I,*}V_{\alpha1},  <-\frac{i}{\sqrt{2}}g_2Z_{mi}^{R,*}V_{\alpha1}>,\nonumber\\
&&\hspace{3cm}H^{S^{\ast}_1{l_j}\overline{F}}_L(1)= -\frac{i}{\sqrt{2}}g_2V_{\alpha1}^{*}Z_{nj}^{R,*},
<-\frac{1}{\sqrt{2}}g_2V_{\alpha1}^{*}Z_{nj}^{I,*}>,\nonumber\\
&&\hspace{3cm}H^{S^{\ast}_1{l_j}\overline{F}}_R(1)= \frac{i}{\sqrt{2}}U_{\alpha2}Z_{nj}^{R,*}Y_{e,j}^{*},
<\frac{1}{\sqrt{2}}U_{\alpha2}Z_{nj}^{I,*}Y_{e,j}^{*}>,\nonumber\\
&&H^{Z{S_1}S^{\ast}_2}(1)=\frac{1}{\sqrt{2}}\Big((g_1\cos\theta_W^\prime\sin\theta_W+g_2\cos\theta_W\cos\theta_W^\prime-g_{YX}\sin\theta_W^\prime)
\sum_{a=1}^3Z_{m,a}^{I,*}Z_{n,a}^{R,*}\nonumber\\&&\hspace{2.2cm}+(g_X\sin\theta_W^\prime)\sum_{a=1}^3Z_{m,3+a}^{I,*}Z_{n,3+a}^{R,*}\Big)(-p_{\mu}^{{\nu}^R_{(n,m)}}+p_{\mu}^{{\nu}^I_{(m,n)}}).
\end{eqnarray}

 Fig.\ref{N1}(2):~$S_1=\widetilde{L}_n$, $S_2=\widetilde{L}_m$, $F=\chi^0$
\begin{eqnarray}\label{A2}
&&H^{S_2F\overline{l}_i}_L(2)= i\Big(-\frac{1}{\sqrt{2}}2g_1N_{\rho1}^{*}Z_{m,3+i}^{E,*}-\frac{1}{\sqrt{2}}(2g_{YX}+g_X)N_{\rho5}^{*}Z_{m,3+i}^{E,*}-N_{\rho3}^{*}Z_{m,i}^{E,*}Y_{e,i}\Big),\nonumber\\
&&H^{S_2F\overline{l}_i}_R(2)= i\Big(\frac{1}{\sqrt{2}}Z_{m,i}^{E,*}(g_1N_{\rho1}+g_2N_{\rho2}+g_{YX}N_{\rho5})-Y_{e,i}^{*}Z_{m,3+i}^{E,*}N_{\rho3}\Big),\nonumber\\
&&H^{S^{\ast}_1{l_j}\overline{F}}_L(2)= i(\frac{1}{\sqrt{2}}g_1N_{\rho1}^{*}Z_{n,j}^E+\frac{1}{\sqrt{2}}g_2N_{\rho2}^{*}Z_{n,j}^E+\frac{1}{\sqrt{2}}g_{YX}N_{\rho5}^{*}Z_{n,j}^E-N_{\rho3}^{*}Y_{e,j}Z_{n,3+j}^E),\nonumber\\
&&H^{S^{\ast}_1{l_j}\overline{F}}_R(2)= i\Big(-\frac{1}{\sqrt{2}}Z_{n,3+j}^E(2g_1N_{\rho1}+(2g_{YX}+g_X)N_{\rho5})-Y_{e,j}^{*}Z_{n,j}^EN_{\rho3}\Big),\nonumber\\
&&H^{Z{S_1}S^{\ast}_2}(2)=\frac{i}{\sqrt{2}}\Big((-g_1\cos\theta_W^\prime\sin\theta_W+g_2\cos\theta_W\cos\theta_W^\prime+g_{YX}\sin\theta_W^\prime)
\sum_{a=1}^3Z_{n,a}^{E,*}Z_{m,a}^E\nonumber\\&&\hspace{2.2cm}+(-2g_1\cos\theta_W^\prime\sin\theta_W+2g_{YX}+2g_X)\sin\theta_W^\prime\sum_{a=1}^3Z_{n,3+a}^{E,*}Z_{m,3+a}^E\Big).
\end{eqnarray}

 Fig.\ref{N1}(3):~$S_1=H^{\pm}$, $S_2=H^{\pm}$, $F=\nu$
\begin{eqnarray}\label{A3}
&&\hspace{2cm}H^{S_2F\overline{l}_i}_L(3)= iU_{ki}^{V,*}Y_{e,i}Z_{m1}^+,~~~~~ H^{S^{\ast}_1{l_j}\overline{F}}_R(3)=iY_{e,j}^{*}U_{kj}^VZ_{n1}^+,\nonumber\\
&&\hspace{5cm}H^{S_2F\overline{l}_i}_R(3)=H^{S^{\ast}_1{l_j}\overline{F}}_L(3)= 0,\nonumber\\
&&H^{Z{S_1}S^{\ast}_2}(3)=\frac{i}{\sqrt{2}}\delta_{ij}\Big(-g_1\cos\theta_W^\prime\sin\theta_W+g_2\cos\theta_W\cos\theta_W^\prime+(g_{YX}+g_X)\sin\theta_W^\prime\Big).
\end{eqnarray}

 Fig.\ref{N1}(4):~$F_1=\chi^c_n$, $F_2=\chi^c_m$, $S=\widetilde{\nu}$
\begin{eqnarray}\label{A4}
&&\hspace{2cm}H^{SF_2\overline{l}_i}_L(4)= -\frac{1}{\sqrt{2}}U_{m2}^{*}Z_{pi}^{I,*}Y_{e,i}, <\frac{i}{\sqrt{2}}U_{m2}^{*}Z_{pi}^{R,*}Y_{e,i}>,\nonumber\\
&&\hspace{2cm}H^{SF_2\overline{l}_i}_R(4)= \frac{1}{\sqrt{2}}g_2Z_{pi}^{I,*}V_{m1}, <-\frac{i}{\sqrt{2}}g_2Z_{pi}^{R,*}V_{m1}>,\nonumber\\
&&\hspace{2cm}H^{S^{\ast}{l_j}\overline{F}_1}_L(4)= -\frac{1}{\sqrt{2}}g_2Z_{pj}^{I,*}V_{n1}^{*}, <-\frac{i}{\sqrt{2}}g_2Z_{pj}^{R,*}V_{n1}^{*}>,\nonumber\\
&&\hspace{2cm}H^{S^{\ast}{l_j}\overline{F}_1}_R(4)= \frac{1}{\sqrt{2}}Z_{pj}^{I,*}Y_{e,j}^{*}U_{n2},
<\frac{i}{\sqrt{2}}Z_{pj}^{R,*}Y_{e,j}^{*}U_{n2}>,\nonumber\\
&&H^{Z{F_1}\overline{F}_2}_L(4)=\frac{i}{\sqrt{2}}\Big(2g_2U_{m1}^{*}\cos\theta_W\cos\theta_W^\prime U_{n1}+U_{m2}^{*}(-g_1\cos\theta_W^\prime\sin\theta_W\nonumber\\
&&\hspace{2.2cm}+g_2\cos\theta_W\cos\theta_W^\prime+(g_{YX}+g_X)\sin\theta_W^\prime)U_{n2}\Big),\nonumber\\
&&H^{Z{F_1}\overline{F}_2}_R(4)=\frac{i}{\sqrt{2}}\Big(2g_2U_{n1}^{*}\cos\theta_W\cos\theta_W^\prime U_{m1}+U_{n2}^{*}(-g_1\cos\theta_W^\prime\sin\theta_W\nonumber\\
&&\hspace{2.2cm}+g_2\cos\theta_W\cos\theta_W^\prime+(g_{YX}+g_X)\sin\theta_W^\prime)U_{m2}\Big).
\end{eqnarray}

 Fig.\ref{N1}(5):~$F_1=\chi^0_n$, $F_2=\chi^0_m$, $S=\widetilde{L}$
\begin{eqnarray}\label{A5}
&&H^{SF_2\overline{l}_i}_L(5)
i\Big(-\frac{1}{\sqrt{2}}2g_1N_{m1}^{*}Z_{k,3+i}^{E,*}-\frac{1}{\sqrt{2}}(2g_{YX}+g_X)N_{m5}^{*}Z_{k,3+i}^{E,*}-N_{m3}^{*}Z_{k,i}^{E,*}Y_{e,i}\Big),\nonumber\\
&&H^{SF_2\overline{l}_i}_R(5)= i\Big(\frac{1}{\sqrt{2}}Z_{k,i}^{E,*}(g_1N_{m1}+g_2N_{m2}+g_{YX}N_{m5})-Y_{e,i}^{*}Z_{k,3+i}^{E,*}N_{m3}\Big),\nonumber\\
&&H^{S^{\ast}{l_j}\overline{F}_1}_L(5)= i\Big(\frac{1}{\sqrt{2}}g_1N_{n1}^{*}Z_{k,j}^E+\frac{1}{\sqrt{2}}g_2N_{n2}^{*}Z_{k,j}^E+\frac{1}{\sqrt{2}}g_{YX}N_{n5}^{*}Z_{k,j}^E-N_{n3}^{*}Y_{e,j}Z_{k,3+j}^E\Big),\nonumber\\
&&H^{S^{\ast}{l_j}\overline{F}_1}_R(5)= i\Big(-\frac{1}{\sqrt{2}}Z_{k,3+j}^E(2g_1N_{n1}+(2g_{YX}+g_X)N_{n5})-Y_{e,j}^{*}Z_{k,j}^EN_{n3}\Big),\nonumber\\
&&H^{Z{F_1}\overline{F}_2}_L(5)=-\frac{i}{\sqrt{2}}\Big(N_{m3}^{*}(g_1\cos\theta_W^\prime\sin\theta_W+g_2\cos\theta_W\cos\theta_W^\prime-(g_{YX}+g_X)\sin\theta_W^\prime)N_{n3}\nonumber\\
&&\hspace{2.2cm}N_{m4}^{*}(g_1\cos\theta_W^\prime\sin\theta_W+g_2\cos\theta_W\cos\theta_W^\prime-(g_{YX}+g_X)\sin\theta_W^\prime)N_{n4}\nonumber\\
&&\hspace{2.2cm}2(-g_X\sin\theta_W^\prime)(N_{m6}^{*}N_{n6}-N_{m7}^{*}N_{n7})\Big){\gamma}_{\mu},\nonumber\\
&&H^{Z{F_1}\overline{F}_2}_R(5)=\frac{i}{\sqrt{2}}\Big(N_{n3}^{*}(g_1\cos\theta_W^\prime\sin\theta_W+g_2\cos\theta_W\cos\theta_W^\prime-(g_{YX}+g_X)\sin\theta_W^\prime)N_{m3}\nonumber\\
&&\hspace{2.2cm}N_{n4}^{*}(g_1\cos\theta_W^\prime\sin\theta_W+g_2\cos\theta_W\cos\theta_W^\prime-(g_{YX}+g_X)\sin\theta_W^\prime)N_{m4}\nonumber\\
&&\hspace{2.2cm}2(-g_X\sin\theta_W^\prime)(N_{m6}^{*}N_{n6}-N_{m7}^{*}N_{n7})\Big){\gamma}_{\mu}.
\end{eqnarray}

 Fig.\ref{N1}(6):~$F_1=\nu_n$, $F_2=\nu_m$, $S=H^{\pm}$
\begin{eqnarray}\label{A6}
&&\hspace{2cm}H^{SF_2\overline{l}_i}_L(6)= iU_{mi}^{V,*}Y_{e,i}Z_{g1}^+,~~~~~H^{S^{\ast}{l_j}\overline{F}_1}_R(6)= iY_{e,j}^{*}U_{nj}^VZ_{g1}^+,\nonumber\\
&&\hspace{5cm}H^{SF_2\overline{l}_i}_R(6)=H^{S^{\ast}{l_j}\overline{F}_1}_L(6)= 0,\nonumber\\
&&H^{Z{F_1}\overline{F}_2}_L(6)=-\frac{i}{\sqrt{2}}\Big((g_1\cos\theta_W^\prime\sin\theta_W+g_2\cos\theta_W\cos\theta_W^\prime-g_{YX}\sin\theta_W^\prime)
\sum_{a=1}^3U_{m,a}^{V,*}U_{n,a}^V\nonumber\\&&\hspace{2.2cm}+(-g_X\sin\theta_W^\prime)\sum_{a=1}^3U_{m,3+a}^{V,*}U_{m,3+a}^V\Big),\nonumber\\
&&H^{Z{F_1}\overline{F}_2}_R(6)=\frac{i}{\sqrt{2}}\Big((g_1\cos\theta_W^\prime\sin\theta_W+g_2\cos\theta_W\cos\theta_W^\prime-g_{YX}\sin\theta_W^\prime)
\sum_{a=1}^3U_{n,a}^{V,*}U_{m,a}^V\nonumber\\&&\hspace{2.2cm}+(-g_X\sin\theta_W^\prime)\sum_{a=1}^3U_{n,3+a}^{V,*}U_{m,3+a}^V\Big).
\end{eqnarray}

 Fig.\ref{N1}(7):~$W_1$, $W_2$, $F=\nu$
\begin{eqnarray}\label{A7}
&&H^{W_2F\overline{l}_i}_L(7)= -\frac{i}{\sqrt{2}}g_2\sum_{a=1}^3U_{ka}^{V,*}{\gamma}_{\mu},~~~~~H^{W^{\ast}_1{l_j}\overline{F}}_L(7)= -\frac{i}{\sqrt{2}}g_2\sum_{a=1}^3U_{ka}^V{\gamma}_{\mu},\nonumber\\
&&\hspace{4cm}H^{W_2F\overline{l}_i}_R(7)=H^{W^{\ast}_1{l_j}\overline{F}}_R(7)= 0,\nonumber\\
&&H^{Z{W_1}W^{\ast}_2}(7)=-ig_2cos\theta_W\cos\theta_W^\prime\Big(g_{\rho\mu}(-p_{\sigma}^{Z_{\mu}}+p_{\sigma}^{W^+_{\rho}})+g_{\rho\sigma}(-p_{\mu}^{W^+_{\rho}}+p_{\mu}^{W^-_{\sigma}})\nonumber\\
&&\hspace{2.2cm}+g_{\sigma\mu}(-p_{\rho}^{W^-_{\sigma}}+p_{\rho}^{Z_{\mu}})\Big).
\end{eqnarray}

 Fig.\ref{N1}(8):~$F_1=\nu_n$, $F_2=\nu_m$, $W$
\begin{eqnarray}\label{A8}
&&\hspace{1cm}H^{WF_2\overline{l}_i}_L(8)= -\frac{i}{\sqrt{2}}g_2\sum_{a=1}^3U_{ma}^{V,*}({\gamma}_{\mu}P_L),~~~~~H^{\overline{F}_1{l_j}W^{\ast}}_L(8)= -\frac{i}{\sqrt{2}}g_2\sum_{a=1}^3U_{na}^V({\gamma}_{\mu}P_L),\nonumber\\
&&\hspace{5cm}H^{WF_2\overline{l}_i}_R(8)=H^{\overline{F}_1{l_j}W^{\ast}}_R(8)= 0,\nonumber\\
&&H^{Z{F_1}\overline{F}_2}_L(8)=-\frac{i}{\sqrt{2}}\Big((g_1\cos\theta_W^\prime\sin\theta_W+g_2\cos\theta_W\cos\theta_W^\prime-g_{YX}\sin\theta_W^\prime)
\sum_{a=1}^3U_{m,a}^{V,*}U_{n,a}^V\nonumber\\&&\hspace{2.2cm}+(-g_X\sin\theta_W^\prime)\sum_{a=1}^3U_{m,3+a}^{V,*}U_{m,3+a}^V\Big),\nonumber\\
&&H^{Z{F_1}\overline{F}_2}_R(8)=\frac{i}{\sqrt{2}}\Big((g_1\cos\theta_W^\prime\sin\theta_W+g_2\cos\theta_W\cos\theta_W^\prime-g_{YX}\sin\theta_W^\prime)
\sum_{a=1}^3U_{n,a}^{V,*}U_{m,a}^V\nonumber\\&&\hspace{2.2cm}+(-g_X\sin\theta_W^\prime)\sum_{a=1}^3U_{n,3+a}^{V,*}U_{m,3+a}^V\Big).
\end{eqnarray}

\end{document}